\documentclass{aa}  
\usepackage{caption}
\usepackage{txfonts}
\usepackage{subfig}
\usepackage{orcidlink}
\usepackage{xparse}
\usepackage{natbib}

\newcommand{\arepo}{{\sc Arepo}}
\newcommand{\smuggle}{\textit{SMUGGLE}}

\newcommand{\mwrlittle}{gal\_r05}
\newcommand{\mw}{gal}
\newcommand{\mwrbig}{gal\_r2}
\newcommand{\mwmbig}{gal\_m2}
\newcommand{\mwmbigrlittle}{gal\_m2r05}
\newcommand{\mwmbigrbig}{gal\_m2r2}
\newcommand{\mwmlittle}{gal\_m05}
\newcommand{\mwmlittlerlittle}{gal\_m05r05}
\newcommand{\mwmlittlerbig}{gal\_m05r2}

\newcommand{\gsim}{\,\lower.7ex\hbox{$\;\stackrel{\textstyle>}{\sim}\;$}}
\newcommand{\lsim}{\,\lower.7ex\hbox{$\;\stackrel{\textstyle<}{\sim}\;$}}

\hypersetup{colorlinks=true,linkcolor=[rgb]{1.,0.2,0.2},citecolor=[rgb]{0.1,0.4,1.},filecolor=[rgb]{0.7,0.2,0.2},urlcolor=[rgb]{0.7,0.2,0.2}}

\begin{document} 

   \titlerunning{Baryon cycle in Milky Way-like galaxies}

   \title{Understanding the Baryon Cycle: Fueling Star Formation via Inflows in Milky Way-like Galaxies}

   \author{Filippo Barbani
          \inst{1, 2}\orcidlink{0000-0002-1620-2577}
          \and
          Raffaele Pascale\inst{2}\orcidlink{0000-0002-6389-6268}
          \and
          Federico Marinacci\inst{1,2}\orcidlink{0000-0003-3816-7028}
          \and
          Paul Torrey\inst{3}\orcidlink{0000-0002-5653-0786}
          \and
          Laura V. Sales\inst{4}\orcidlink{0000-0002-3790-720X}
          \and
          Hui Li\inst{5}\orcidlink{0000-0002-1253-2763}
          \and
          Mark Vogelsberger\inst{6, 7}\orcidlink{0000-0001-8593-7692}
          }

   \institute{Department of Physics and Astronomy "Augusto Righi", University of Bologna, Via P. Gobetti 93/2, I-40129 Bologna, Italy\\
              \email{filippo.barbani2@unibo.it}
         \and
             INAF, Astrophysics and Space Science Observatory Bologna, Via P. Gobetti 93/3, I-40129 Bologna, Italy
             \and
             Department of Astronomy, University of Virginia, Charlottesville, VA 22904, USA
             \and
             Department of Physics and Astronomy, University of California, Riverside, CA 92521, USA
             \and
             Department of Astronomy, Tsinghua University, Beijing 100084, China
             \and
            Department of Physics and Kavli Institute for Astrophysics and Space Research, Massachusetts Institute of Technology, Cambridge, MA 02139, USA
             \and
            The NSF AI Institute for Artificial Intelligence and Fundamental Interactions, Massachusetts Institute of Technology, Cambridge, MA 02139, USA
             }

   \date{Received xxx; accepted xxx}

 
  \abstract
   {Galaxies are not isolated systems; they interact with their surroundings throughout their lifetimes by both ejecting gas via stellar feedback and accreting gas from their environment. Understanding the interplay between the gas ejected from the disc and the circumgalactic medium (CGM) is crucial to learning how star-forming galaxies evolve.}
   {Our goal is to understand how gas in the CGM is accreted onto the inner regions of the star-forming disc, making it available for the formation of new stars. Specifically, we explore the connection between stellar feedback and gas accretion from the CGM in Milky Way-like galaxies, aiming to unveil the complex mechanisms driving the evolution of star-forming galaxies. We focus on the distribution of vertical and radial gas flows to and from the disc as a function of galactocentric radius, and examine the implications of these processes for the evolution of such galaxies.}
   {We use the moving-mesh code \arepo\ coupled with the \smuggle\ sub-grid  model to perform hydrodynamic $N$-body simulations of 9 different galaxies surrounded by a hot ($T \sim 10^6\,{\rm K}$) CGM (also called galactic corona). Each simulation has a different structure of the gaseous disc in terms of mass and scale length, which allows us to study how the dynamics of the gas can be affected by disc structure.}
   {We find evidence of a crucial link between stellar feedback processes and gas accretion from the CGM, which together play an essential role in sustaining ongoing star formation in the disc. In particular, the ejection of gas from the plane of the disc by stellar feedback leads to the generation of a baryon cycle in which the CGM gas is preferentially accreted onto the external regions  of the disc ($ \approx 3-10$ M$_{\odot}$ yr$^{-1}$ of gas is accreted into the entire disc). From these regions it is then transported to the centre with radial mass rates $\approx 1-4$ M$_{\odot}$ yr$^{-1}$ on average, owing to angular momentum conservation, forming new stars and starting the whole cycle again. We find that both vertical accretion onto the inner regions of the disc and the radial transport of gas from the disc outskirts are necessary to sustain star formation.}
   {}

   \maketitle
%

\section{Introduction}
  Observations show that Milky Way-like galaxies have formed stars at an almost constant star formation rate (SFR) over the past gigayears \citep{binney2000,wyse2009,madau2014, haywood2014, gondoin2023}. Given the fact that the gas depletion time in these galaxies is $\approx 1-2$ Gyr \citep[][]{saintonge2013, tacconi2018}, this would not be possible without a substantial accretion of gas onto the star-forming disc over time. Also, the observed metallicities of stars in the solar neighborhood require a consistent low metallicity source of accretion onto the galactic disc, which is known as the G-dwarf problem \citep[e.g.][]{vandenbergh1962,alibes2001}. Thus, accretion of gas from surrounding environment is crucial to understand galaxy formation and evolution. 

Numerical simulations are a fundamental instrument in advancing our understanding of galaxy formation and evolution \citep[see][and references therein]{vogelsberger2020}, offering key insights into the mechanisms driving gas inflows and star formation. Results from simulations suggest that gas accretion occurs primarily in two main modes: cold and hot accretion. In the cold accretion mode \citep{keres2005,dekel2009,vandevoort2011,nelson2016}, that is typically dominant at high redshift ($z\gsim 2$), cold filaments from the intergalactic medium can survive their journey through the galaxy halo and deliver cold gas directly into the star-forming disc. On the other hand, in the hot accretion mode \citep{dekel2006,keres2009,nelson2013,stern2020}  the gas that is falling into the dark matter potential well is shock-heated to the virial temperature, forming a halo of hot gas that subsequently must cool before being accreted onto the galaxy. This mode is generally prevailing at lower redshifts ($z\lsim 2$) for galaxies like the Milky Way and might be fundamental to the build up of discs \citep{Sales2012,Ubler2014,Hafen2022,Yu2023}.

This picture becomes more complex over gigayears of evolution, as the environment around the galactic disc develops a complex, multi-phase structure, generating the so-called circumgalactic medium (CGM). The CGM is confirmed to be a multiphase medium by observations \citep[e.g.][]{tumlinson2011, putman2012, anderson2016, prochaska2017}, containing cold ($T\lesssim10^4$ K), warm ($10^4 \lesssim T\lesssim10^6$ K) and hot ($T \gtrsim 10^6$ K) gas. Furthermore, this component of the galactic ecosystem contains a significant fraction of the baryons associated to Milky Way-like galaxies (up to 50\%, see \citealt{werk2014,bregman2018}). This makes the CGM the primary reservoir from which star-forming galaxies can get the fresh gas they need to sustain star formation.
Hot gas in the CGM has been observed mainly through X-ray emission both in external galaxies \citep{anderson2011,walker2015,zhangyi2024,locatelli2024} and in the Milky Way \citep{yoshino2009,bluem2022,Bhattacharyya2023} and also in the Large Magellanic Cloud \citep{gulick2021}. 

Despite these observations, one problem that is still not fully understood is how the hot gas in the CGM is cooling to accrete into the galactic disc. Although gas can cool through the spontaneous growth of thermal instabilities, this scenario has been ruled out by linear perturbation analysis \citep[see, e.g.,][]{binney2009,nipoti2010} but could be still relevant in case of a substantial rotation of the gas in the inner regions of the corona \citep{sormani2019}. Another interesting proposal is that gas may cool thanks to the mixing with the galactic fountains, gas clouds of mainly cold gas that are ejected from star-forming regions (i.e. the disc) by stellar feedback processes. These gas flows travelling through the CGM can let a fraction of the hot gas condensate into the wake of the cold fountain clouds, increasing the amount of gas that then falls back onto the disc, therefore feeding the galaxy with low metallicity gas that can be used to form new stars. This scenario has gained support from hydrodynamical simulations both at parsec-scale \citep{marinacci2010,marinacci2011,armillotta2016} and galactic-scale \citep{hobbs2020, barbani2023}.

Stellar feedback processes, such as supernova explosions, play a crucial role in shaping galaxy evolution and are responsible for launching galactic outflows that expel metal-rich gas from the galactic disc. These include galactic winds \citep{veilleux2005}, strong outflows that can carry gas out of the galaxy entirely, and galactic fountains \citep{shapiro1976}, which recirculate gas closer to the disc. The latter, in particular, are a primary focus of this work. The processes that involve the ejection of gas from the galactic disc and the subsequent accretion are generating the so-called baryon cycle. In particular, discovering the spatial distribution of the gas accretion and of the galactic fountains is crucial for the evolution of Milky Way-like galaxies. Studying gas accretion and outflows in external galaxies is notoriously difficult, but in the past years, the distribution of gas accretion has been analysed in observations \citep{zheng2017, bish2019,li2023,zhangyi2024, chastenet2024} and studied in numerical simulations \citep{Ubler2014,nuza2019,iza2021,trapp2022,iza2024, roy2024}.

Accretion of gas is expected to happen
predominantly in the outer regions of the disc due to angular momentum conservation and to a larger cross-section of the external disc regions, therefore a mechanism that move the accreted gas to the centre of the galaxy is needed to keep forming new stars. \citet{lacey1985} proposed three mechanisms that could generate these radial flows: if the specific angular momentum of the infalling gas is lower than the one of the gas in the disc we expect a radial flow towards the centre. Other mechanisms proposed by \citet{lacey1985} are the presence of viscosity in the gas layer and gravitational interaction between the gas in the disc and spiral arms and bars. The \ion{H}{i} 21 cm emission line is the best way to investigate these radial motions in the outskirts of spiral galaxies, but it is still unclear if the radial inflows are sufficient to bring enough gas to the centre to sustain the star formation \citep{wong2004, schmidt2016, diteodoro2021}. Finally, these radial flows are necessary also to reproduce the metallicity gradients \citep[e.g.][]{mott2013} and the exponential density profiles \citep[e.g.][]{wang2014} observed in spiral galaxies.

Understanding how CGM is accreted into the galactic disc is still puzzling. The goal of this work is to understand how the CGM can become available for the formation of new stars in the centre of the disc, therefore we investigate the distribution of vertical and radial outflow and inflows over the disc and how they are connected to the structure of the disc itself. To achieve this goal, we perform high-resolution hydrodynamical $N$-body simulations with the \textit{S}tars and \textit{MU}ltiphase \textit{G}as in \textit{G}a\textit{L}axi\textit{E}s (\smuggle) model, an explicit ISM and stellar feedback model for the moving-mesh code \arepo. We simulate a set of 9 star-forming Milky Way-like galaxies including a hot gaseous halo surrounding the galactic discs. We vary the gas disc mass and size, in order to quantify the key properties of vertical and radial gas inflows/outflows and investigate their connection to the evolution of the whole galactic disc. The paper is organised as follows. In Section \ref{sec2} we describe the numerical methods and how we built the initial conditions for the simulations. In Section \ref{result} we present our main findings, comparing them with previous works in Section \ref{comparison}. Finally, Section \ref{summary} provides a brief summary of our results.


\begin{table*}
	\centering
	\caption{Structural parameters of the fiducial Milky Way-like galaxy simulated in this work.}\label{tabella_param}
	\begin{tabular}[t]{ccccccccccccc}
	
		\hline
        \hline
		$R_{200}$&$M_{\text{dm}}$&$r_{\text{s}}$&$M_{\text{b}}$ & a & $M_{\star}$&$h_{\star}$& $z_{\star}$ &$M_{\text{g}}$ & $h_{\text{g}}$ & $M_{\text{cor}}$&$r_{\text{c}}$ &$\beta$\\

		[kpc]&$[M_{\odot}]$&[kpc]&$[M_{\odot}]$&[kpc]&$[M_{\odot}]$&[kpc] & [pc] & [$M_{\odot}$] & [kpc] & [$M_{\odot}$]&[kpc]& \\[0.15cm]
		\hline
		$241$ & $1.53\times 10^{12}$&36.46&$1.5\times 10^{10}$&1.3&$4.73\times 10^{10}$&3.8 & 380 & $9\times 10^{9}$ & 7.6 & $5\times 10^{10}$ &83&2/3\\
		
		\hline
        \hline
	\end{tabular}
    \tablefoot{From left to right: virial radius $R_{200}$; dark matter halo mass ($M_{\text{dm}}$); dark matter halo scale length ($r_{\text{s}}$); bulge mass ($M_{\text{b}}$); bulge scale length ($a$); stellar disc mass ($M_{\star}$); stellar disc scale length ($h_{\star}$); stellar disc scale height ($z_{\star}$); gaseous disc mass ($M_{\text{g}}$); gaseous disc scale length ($h_{\text{g}}$); mass of the corona ($M_{\text{cor}}$) computed within the virial radius $R_{200}$; corona core radius ($r_{\text{c}}$); beta model parameter ($\beta$).}
\end{table*}%

\section{Numerical methods}\label{sec2}
The simulations in this work are performed with the moving mesh code \arepo\ \citep{springel2010, pakmor2016, weinberger2020}, a hydrodynamical $N$-body code that solves the equations of hydrodynamics on an unstructured Voronoi mesh able to move with the flow of the gas. On this moving mesh, \arepo\ integrates Euler equations with a finite-volume Godunov method that employs an exact Riemann solver. Gravitational accelerations are computed with an oct-tree method \citep{barnes1986} and gravitational dynamics is evolved with a leapfrog algorithm. The code also uses a mesh refinement scheme, ensuring that each gas cell keeps its mass within a factor of two from a predetermined target value; together with the moving nature of the mesh this guarantees an automatic adjustment of the resolution in regions of high density. More details on \arepo\ can be found in \citet{springel2010} and \citet{weinberger2020}.

\subsection{The \smuggle\ model}

Coupled with the \arepo\ code we use the \smuggle\ model, a sub-grid model which incorporates baryonic processes that are fundamental for the evolution of star-forming galaxies. The major processes implemented by the model are cooling and heating of the gas, star formation and evolution, and stellar feedback. We briefly describe these key constitutive elements of \smuggle\ below and we refer the reader to \citet{marinacci2019} for a complete description of the model.

\smuggle\ includes primordial and metal cooling \citep{vogelsberger2013}. Cooling rates are tabulated as a function of density, temperature and redshift using {\sc cloudy} \citep{ferland1998} and considering the presence of a UV background \citep{faucher2009}. An important feature of the model, which allows the gas to reach high densities and form new stars, is the generation of a low-temperature ISM phase ($T\approx 10$ K) due the inclusion of low-temperature metal lines, fine structure and molecular cooling \citep[see][]{Hopkins2018}. This low-temperature cooling is facilitated by the implementation of gas self-shielding from the UV background, which follows the parameterization described in \citep{rahmati2013}. In addition to stellar feedback (see below), gas heating can occur through cosmic rays \citep{guo2008} and photo-electric \citep{wolfire2003} heating, which are important for the thermal balance of $T\approx50$ K and $T\approx8000$ K gas in the ISM.

New stars can be formed in high density ($\rho > \rho_{\text{th}}=100$ cm$^{-3}$) and gravitationally bound gas cells. If a gas cell is eligible for star formation, it forms stars at a rate $\dot{M}_{\star}= \epsilon_{\text{SF}} M_{\text{g}} / t_{\text{ff}}$, where $M_{\text{g}}$ is the gas cell mass, $t_{\text{ff}}$ is the free-fall time and $\epsilon_{\text{SF}}(=0.01)$ is the star formation efficiency per free-fall time. Instead of forming stars in a continuous way, star formation works stochastically, transforming gas cells in star particles with a probability which is compatible with $\dot{M}_{\star}$. These star particle represent a single stellar population that follows a \citet{chabrier2003} initial mass function (IMF). As they evolve, star particles inject mass and metals -- interpolated from a set of stellar evolution tables \citep[see][for details]{vogelsberger2013} -- into ISM, changing its chemical composition. Additionally, they deliver energy and momentum to the interstellar gas through stellar feedback processes, which are implemented by considering the following channels:
\begin{itemize}
\item[(i)] Supernova (SN) feedback is crucial in regulating star formation and the ISM structure through injection of momentum and energy \citep{martizzi2015, kim2018}, while also being capable of ejecting multiphase gas into the CGM \citep{fielding2022}, distributing gas and metals into the galactic halo. To implement type II SN feedback, at each time-step the number of SNe is computed from the adopted \citet{chabrier2003} IMF (considering stars in the mass range $8<M/ M_{\odot}<100$).   Instead, the temporal distribution of type Ia SN events is parameterized with a delay time distribution (\citealt{maoz2012}), from which the number of events is derived. From these rates, the injection of energy, momentum, mass, and metals into the ISM can be computed by assuming an energy of $10^{51}$ erg per SN. One problem in modeling SN feedback, that arises due to insufficient resolution of galaxy scale simulations, is how to properly describe the Sedov-Taylor phase of the SN remnant in which most of the momentum imparted to ISM is generated. To accurately capture the momentum generation, the model explicitly accounts for the momentum boost occurring during the Sedov-Taylor phase with a formulation derived from individual, high-resolution SN simulations \citep[][see also \citealt{Hopkins2018b}]{cioffi1988}.

\item[(ii)] Radiative feedback is produced by massive young stars (stellar particles with an age $\lesssim 5\,{\rm Myr}$) and it is important for the thermal and dynamical state of the ISM. Indeed, several works suggest that it can disrupt giant molecular clouds, therefore making SNe more effective \citep[e.g.][]{walch2012,smith2021}. In \smuggle\ this feedback channel includes photoionization and radiation pressure. Photionization is treated probabilistically and each gas cell surrounding a young star particle has a probability of being photoionized that depends on the ionizing photon rate and the cell recombination rate. If a gas cell is photoionized, its temperature is set to $1.7\times 10^4$ K and its cooling is stopped for a duration equal to the time step of the (photo-)ionizing star particle. The pressure generated from radiation produced by young stars is a source of momentum that is directly injected in the cells surrounding the star particle. 

\item[(iii)] Stellar winds are produced by OB and AGB stars, which contribute to feedback at different times due to the different evolutionary scales of these types of stars \citep[e.g.][]{krumholz2009}. To account for the effect of stellar winds, \smuggle\ computes the mass loss for each star particle at any given time-step. This quantity is subsequently used to derive the energy and the momentum injected into the surrounding gas particles.
\end{itemize}
For all feedback channels, mass, linear momentum, and energy are coupled to the gas in a fully conservative manner, ensuring their conservation.

By including the aforementioned key physical processes, \smuggle\ is able to capture the complex multi-phase structure of the ISM  while simultaneously regulating star formation to match the observed levels in star-forming galaxies. Moreover, the explicit modeling of stellar feedback enables the self-consistent generation of galactic-scale gas flows to and from the disc. Owing to these capabilities, \smuggle\ has been successfully and extensively used to investigate various topics related to the evolution of galaxies and their ISM/CGM. These include, among others, the properties of giant molecular clouds \citep{Li2020}, the formation and evolution of young stellar clusters \citep{li2022}, the development of cored dark matter profiles in dwarf galaxies \citep{Burger2022a,Burger2022b, Jahn2023}, the kinematics of stellar bars in spiral galaxies \citep{Beane2023}, the impact of feedback on the distribution of \ion{O}{vi} absorbers \citep{zhangzhijie2024}, the formation of polycyclic aromatic hydrocarbons \citep{Narayanan2023}, the evolution and distribution of super-bubbles in simulated disc galaxies \citep{Li2024}, and the effects of AGN feedback in galaxies with multiphase ISM \citep{sivasankaran2024}.

\begin{figure}
\centering
\includegraphics[width=\columnwidth]{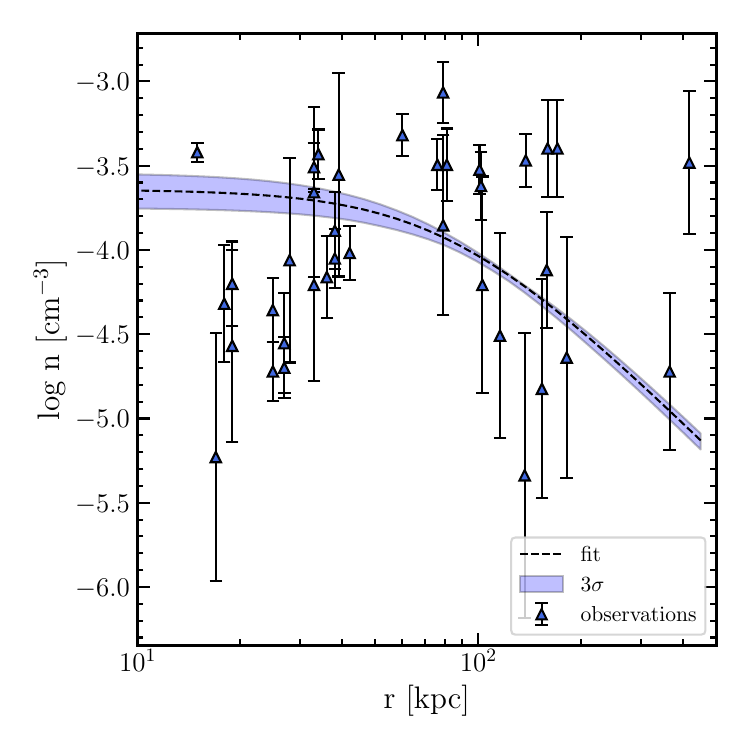}

\caption{Number density of the coronal gas as a function of radius. The black dashed line is the median model obtained by fitting a beta model to Milky Way coronal gas observations (upward pointing triangles with errorbars) using a Markov chain Monte Carlo algorithm. The coloured shaded area represents the 3$\sigma$ region. Observations, which should be considered as lower limits for the coronal density, are taken from \citet{putman2021}.}\label{profile_corona}

\end{figure}

\begin{table*}
	\caption{Summary of model parameter variations.}\label{model_var}
    \centering
	\begin{tabular}[t]{cccccccccc}
	
		\hline
        \hline
		Simulation name & \mwrlittle & \mw & \mwrbig & \mwmlittlerlittle & \mwmlittle & \mwmlittlerbig & \mwmbigrlittle & \mwmbig & \mwmbigrbig\\
        \hline
	    \rule{0pt}{3ex}  
		$f_{\text{M}}$ & 1 & 1 & 1 & 0.5 & 0.5 & 0.5 & 2 & 2 & 2 \\
        \hline
        \rule{0pt}{3ex}  
		$f_{\text{G}}$ & 0.5 & 1 & 2 & 0.5 & 1 & 2 & 0.5 & 1 & 2 \\
		\hline
        \hline
	\end{tabular}
    \tablefoot{Summary of the simulations performed in this paper, each varying specific model parameters relative to the fiducial run (labelled as \mw). The initial conditions differ in the mass of the gaseous disc, scaled by $f_{\text{M}}$ relative to the fiducial value, and its scale length, scaled by $f_{\text{G}}$. The fiducial values for these parameters are provided in Table \ref{tabella_param}.}
\end{table*}

\subsection{Initial conditions}

The generation of initial conditions is based on the method presented in \citet{springel2005}, which enables the creation of a multi-component galaxy in approximate equilibrium \citep[see also][for a description of this method tailored to the type of simulations examined in this work]{barbani2023}. We consider a galaxy including the following components: a dark matter halo and a stellar bulge, a thick stellar disc, a thick gaseous disc and a galactic hot corona (i.e., the hot phase of the galaxy CGM).

The dark matter halo and the bulge are modelled with a Hernquist profile

\begin{equation}\label{profil_hern}
	\rho_{*}(r)=\frac{M_*}{2\pi} \frac{a_{*}}{r(r+a_{*})^3},
\end{equation}
where $*=$ dm for the dark matter halo and $*=$ b for the bulge. $a_*$ is the scale radius and $M_*$ is the total mass. The dark matter is modelled as a static gravitational potential to reduce computational time.

The radial distribution of both stars and gas follow an exponential profile, as in observations \citep{freeman1970}. The surface density of this profile is given by

\begin{equation}\label{star_surf}
 	\Sigma_{*}(R)=\frac{M_*}{2\pi h_{*}^2}\text{exp}(-R/h_{*}) \ ,
 	\end{equation}
 where $*=\star$ for stars and $*=\text{g}$ for gas. $h_*$ is the scale length and $\Sigma_*$ is the surface density of the stellar/gaseous discs. Vertically, the stellar disc has a $\text{sech}^2$ profile
 \begin{equation}\label{sec}
	\rho_{\star}(R,z)= \frac{\Sigma_{\star}(R)}{2 z_{\star} } \text{sech}^2\biggl(\frac{z}{z_{\star}} \biggr),
	\end{equation}
 where $z_{\star}$ is the scale height and $\Sigma_{\star}$ is the surface density of the stellar disc. The gas vertical profile is determined by iteratively solving the equation for vertical hydrostatic equilibrium $\frac{1}{\rho_g} \partial P/ \partial z = -\partial \Phi/ \partial z$, where $\Phi$ is the total gravitational potential and $P$ is the gas pressure. In this way we obtain the gas density $\rho_g$, that is then used to evaluate
 \begin{equation}\label{sigma_int}
	\Sigma_g(R)=\int_{-\infty}^{+\infty} \rho_g(R,z) \, \text{d}z
	\end{equation}
and the process is repeated until the desired gas surface density profile $\Sigma_g(R)$ is obtained.

The metallicity in the gaseous disc is modeled as an exponential profile based on observation of the metallicity gradient in the Milky Way \citep{lemasle2018} with slope = $-0.0447$ dex kpc$^{-1}$ and intercept = $0.3522$ dex, thus the metallicity at the solar radius $R\sim8$ kpc is $1Z_{\odot}$. 

\begin{figure*}
\centering
\includegraphics[width=0.95\textwidth]{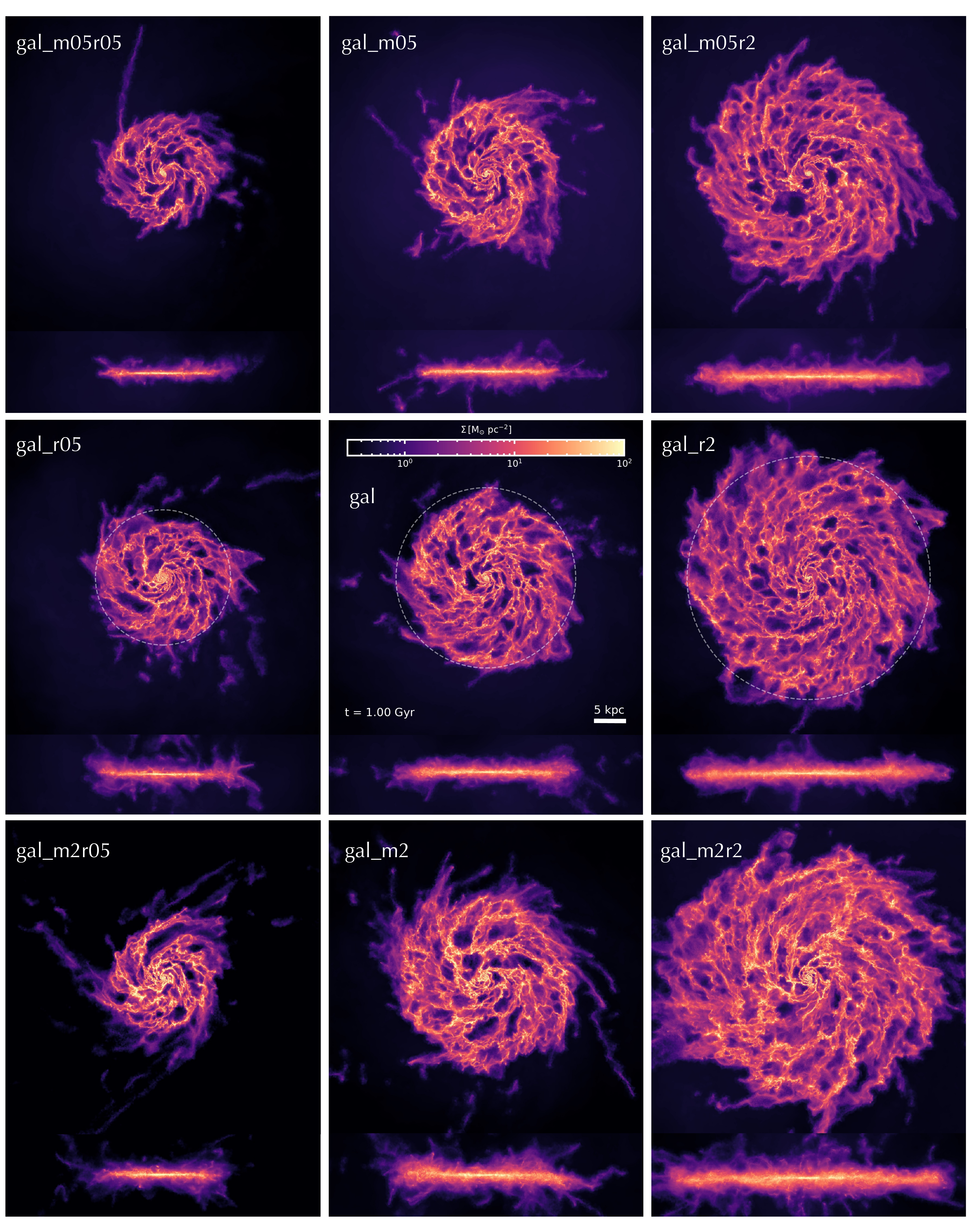}
\caption{Gas column density face-on (top panels of each row) and edge-on (bottom panels of each row) projections computed using the  \smuggle\ model at $t=1$ Gyr for all the 9 simulations examined in this work, as labelled in each panel. The simulations differ in the initial value of the scale-length and of the mass of the gaseous disc (see Table \ref{model_var}). Each panel is 70 kpc across and in projection depth with a total number of 1024 × 1024 pixels that give a resolution of $\sim70$ pc. Brighter colours correspond to higher column densities, as indicated in the colorbar. In the \mwrlittle, \mw \ and \mwrbig \ projection panels, the dashed white circle shows the disc size $R_d$, defined as the radius at which the gas surface density drops below 1 M$_{\odot}$ pc$^{-2}$ (not shown in the other projections for clarity).}\label{density_proj}
\end{figure*}

In order to investigate the interaction between the disc and the surrounding medium, we added the hot component of the CGM -- i.e. galactic corona -- around the galaxy. Here, we briefly explain how this component is included in the initial conditions. The coronal density profile is described by a $\beta$ model of the form
\begin{equation}
    \rho_{\text{cor}}(r) = \rho_0 \biggl[1 + \biggl(\frac{r}{r_c}\biggr)^2\biggr]^{-\frac{3}{2} \beta},
\end{equation}
where $\beta = 2/3$ \citep{jones1984, moster2011}, $\rho_0$ is the central density and $r_c$ is the core radius. Figure \ref{profile_corona} shows the number density profile for the galactic corona. Points with errorbars are observational data for the density of the Milky Way corona \citep{putman2021}, which are obtained from dwarf galaxies observations by evaluating the coronal density necessary to strip the gas of the dwarf at the perigalacticon. Hence, they should be considered as lower limits for the density of the corona. The $\beta$ model profile of the corona has formally infinite mass, but we truncated the resulting density distribution at $10 \,R_{200}$. To obtain the central density $\rho_0$ and the core radius $r_c$ we applied a Markov chain Monte Carlo algorithm using an affine-invariant ensemble sampler ({\sc emcee} package, \citealt{foreman2013}). From the Markov chain Monte Carlo fit we found a core radius $r_c = 83$ kpc and a central density $\rho_0 = 3.3\times 10^3$ M$_{\odot}$ kpc$^{-3}$. The coronal gas particles are initialized with an azimuthal rotational velocity $v_{\phi} = \alpha v_c$, where $v_c=(R\partial\Phi/ \partial R)^{1/2}$ is the circular velocity due to the total gravitational potential $\Phi$. $\alpha$ is set to $0.4$ to ensure a rotational velocity $v_{\phi}\approx90$ km s$^{-1}$ close to the galactic disc \citep{marinacci2011}. Then, the temperature profile is obtained solving an equation analogous to the hydrostatic equilibrium equation using the effective potential $\Phi_{\text{eff}}$
\begin{equation}
 \Phi_{\text{eff}} = \Phi(R, z) - \int_{\infty}^{R} \frac{v_{\phi}^2(R')}{R'} \ \text{d} R'.
\end{equation}
The effective equilibrium of each system has been verified by running adiabatic simulations, i.e. without dissipative processes such as radiative cooling. The metallicity of the coronal gas is set to Z $=0.1$ Z$_{\odot}$, in line with observational estimates \citep[e.g.][]{bogdan2017}. Finally, we note that the profiles adopted to initialize all components of the systems in our simulations are axisymmetric. Therefore, the initial conditions have no azimuthal variations.

Table \ref{tabella_param} lists the structural parameters of our fiducial Milky Way-like galaxy model, the other eight galaxies investigated in this work are variations of this fiducial model. In Table \ref{model_var} we summarise the different parameter variations that we consider in our runs. Specifically, we scale the mass of the gaseous disc $M_{\text{g}}$, by a factor $f_\text{M}$, and the gaseous disc scale length $h_{\text{g}}$, by a factor $f_\text{G}$, with respect to the fiducial values $9\times 10^9\,M_{\odot}$ and $7.6$ kpc -- in practice our simulation set explores variations of a factor of two with respect to these reference values. The gravitational softening length is set at a minimum of $\epsilon_{\text{gas}} = 10$ pc for the gas and at $\epsilon_{\star} = 21.4$ pc for the stars for all the nine runs. In the fiducial run there are $N_{\text{disc}}=3.2\times 10^6$ particles in the stellar disc, $N_{\text{bulge}}=8\times 10^5$ in the bulge and $N_{\text{gas}}\simeq 1.1\times 10^7$ gaseous particles, with $N_{\text{gas,d}}=3.2\times 10^5$ gaseous particles in the disc and $N_{\text{gas,cor}}\simeq1.1\times 10^7$ in the corona. This results in a mass resolution $m_{\text{gas}}=1.1\times10^4$ M$_{\odot}$, $m_{\text{disc}}=1.5\times10^4$ M$_{\odot}$ and $m_{\text{bulge}}=2\times10^4$ M$_{\odot}$ for the stellar disc and bulge particles,  respectively. The number of gas particles in the disc in the simulations where the mass of this component is varied with respect to the reference value is changed accordingly to ensure the same mass resolution. 
All nine galaxies achieve an approximate state of equilibrium within $\sim0.3$ Gyr and are evolved for a total time span of 2 Gyr.

\begin{figure}
\centering
\includegraphics[width=\columnwidth]{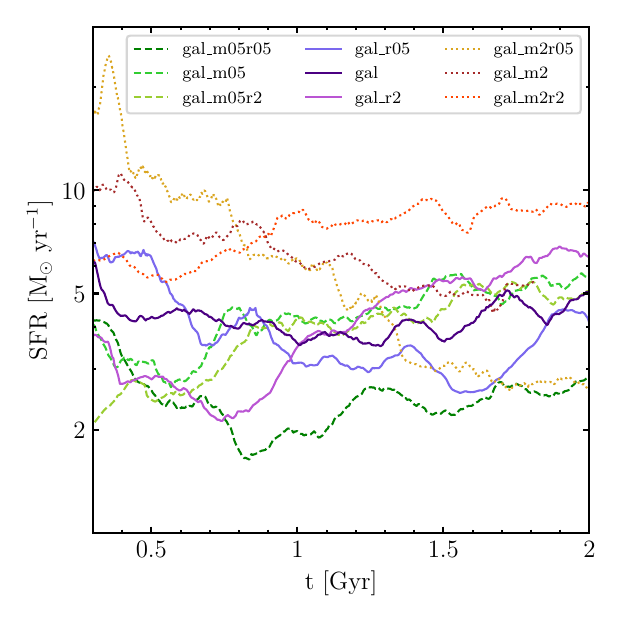}

\caption{Star formation rate as a function of time for the 9 simulations analysed in this work. The SFR has a wide range of values ($\approx 2-10$ M$_{\odot}$ yr$^{-1}$) across the different simulations, but it remains approximately constant over time for each simulated system. The average SFR depends on the disc initial properties, with variations in each simulation staying within a factor of $\sim2$ from the average value.}\label{sfr}
\end{figure}

\begin{figure*}
\centering
\subfloat{\includegraphics[width=\textwidth]{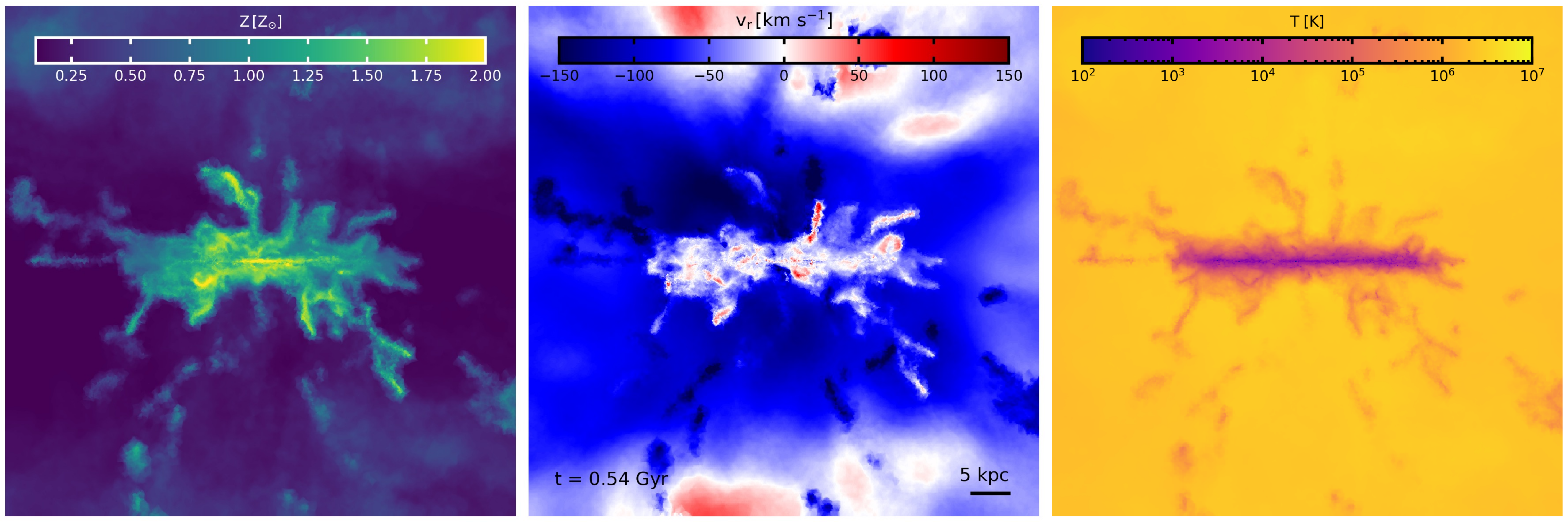}} 

\caption{Density-weighted maps of the gas metallicity, radial velocity and temperature (from left to right) in edge-on projections of the fiducial simulation at $t=0.54$ Gyr. Each panel is 70 kpc across and in projection depth with a total number of 1024 × 1024 pixels that give a resolution of $\sim70$ pc. These projections show the stark difference between the disc and the CGM properties.}\label{metal_edge}

\end{figure*}

\begin{figure*}
\centering
\subfloat{\includegraphics[width=\textwidth]{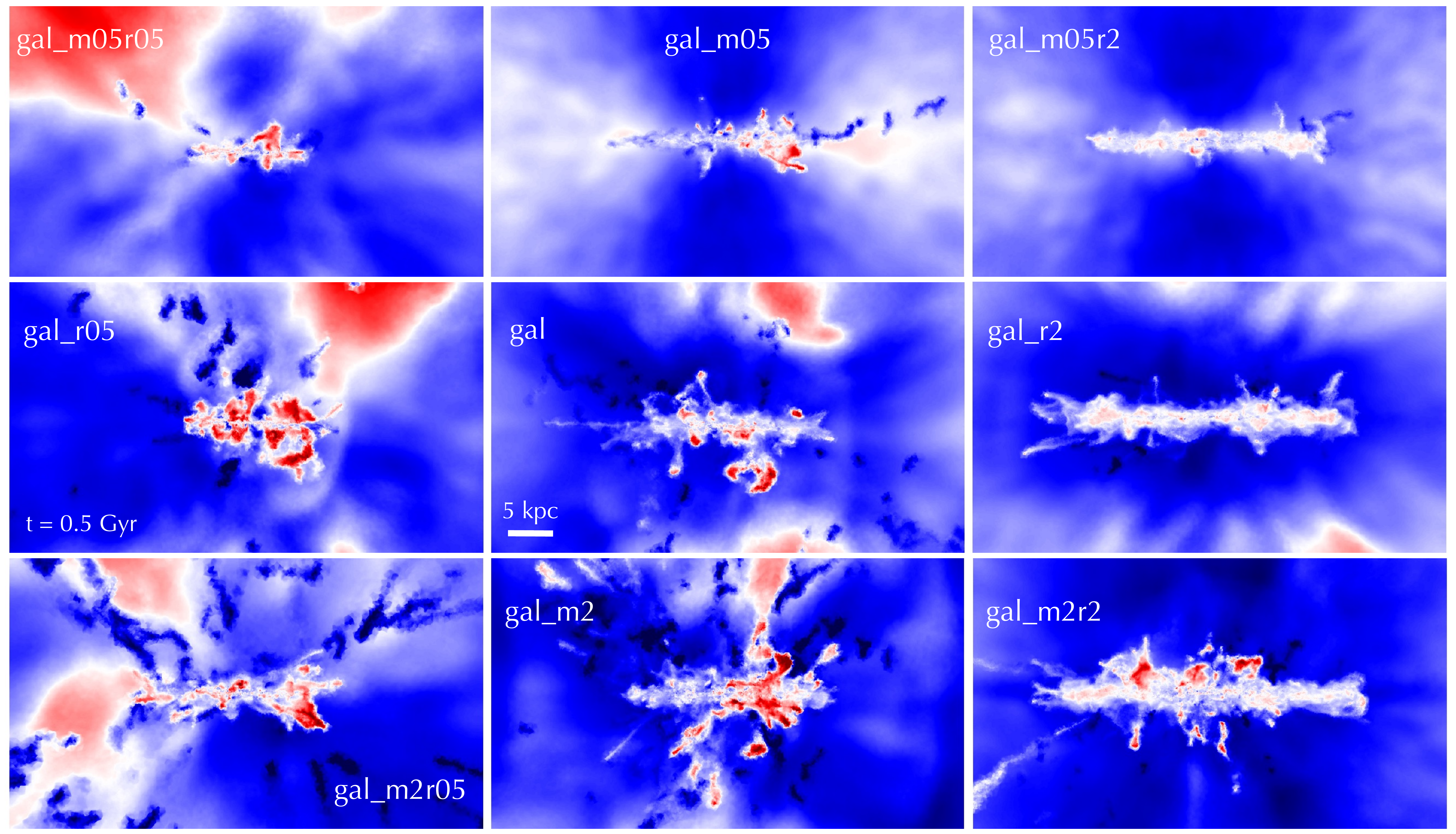}} 

\caption{Density-weighted maps of the gas radial velocity in edge-on projections of the 9 simulations at $t=0.5$ Gyr. Each panel is $70\, {\rm kpc}\times40$ kpc and with a projection depth of 70 kpc with a resolution of $\sim70$ pc. The velocities go from -150 km s$^{-1}$ (inflows, blue) to 150 km s$^{-1}$ (outflows, red) as in central panel of Figure \ref{metal_edge}. These projections clearly show the presence of galactic fountains in the simulations.}\label{proj_radial}

\end{figure*}

\section{Results}\label{result}

We now discuss the main results of our simulations. In Section \ref{galaxystruc} we present the structure of the galaxies and of the gas in our simulations and we analyse the connection to the SFR. In Section \ref{vertical} we focus on the vertical ejection/accretion of gas from/into the galactic disc, studying also the net accretion rate and how it changes between the 9 simulations. In Section \ref{radial} we analyse radial gas motions in the disc plane and their importance for the baryon cycle. Finally, in Section \ref{insideout} we look at the disc size temporal evolution and its link to the vertical outflow/inflow rates.

\subsection{Galaxy and outflow/inflow structure and distribution}\label{galaxystruc}

\begin{figure}
\centering
\includegraphics[width=\columnwidth]{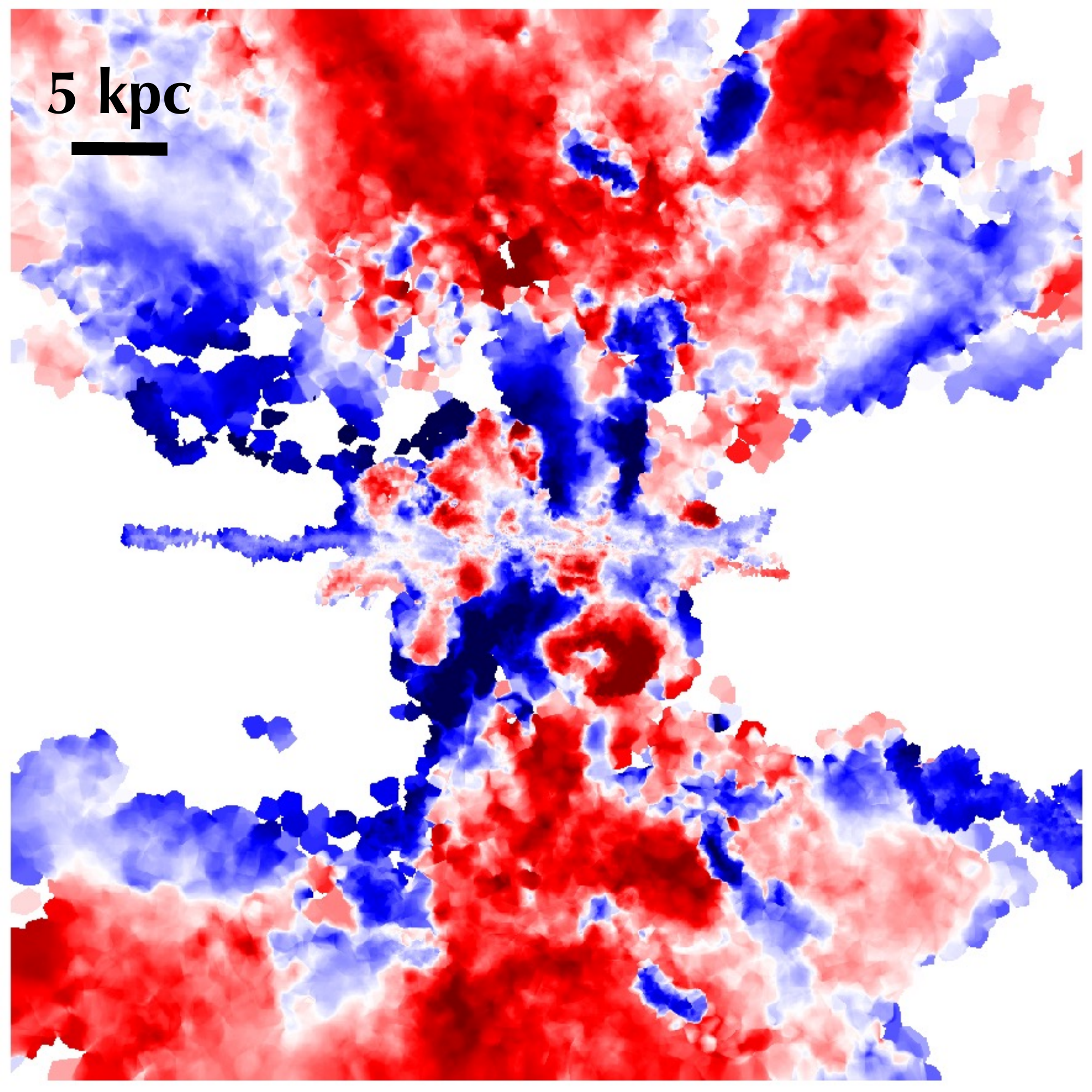}

\caption{Density-weighted maps of the high-metallicity ($Z>0.9 \, Z_{\odot}$) gas radial velocity in edge-on projection of the \mw \ simulation at $t=0.8$ Gyr. The velocities range from -150 km s$^{-1}$ (inflows, blue) to 150 km s$^{-1}$ (outflows, red) as in central panel of Figure \ref{metal_edge}. The projection clearly highlights how the accreted gas is not pristine material coming from the CGM, but is rather a mixture of the latter with gas that was previously ejected from the disc.}\label{high_metal_map}
\end{figure}

We begin our discussion by presenting the general properties and the appearance of the gaseous discs in our runs. Figure \ref{density_proj} shows the gas projected density distribution (edge-on and face-on), after 1 Gyr of evolution, of all 9 simulations considered in this work. Each row differs in the mass of the gaseous disc, whereas in columns only the initial scale length the gaseous disc is varied. The simulations are arranged, from left to right, as follows (see Table \ref{model_var} for their main properties): \mwmlittlerlittle, \mwmlittle \ and \mwmlittlerbig \ (first row), \mwrlittle, \mw \ and \mwrbig \ (second row), \mwmbigrlittle, \mwmbig \ and \mwmbigrbig \ (third row). In the second row, the central panel shows the fiducial galaxy; in the left panel the disc scale length is reduced by a factor $f_\text{G} = 1/2$ relative to the reference case, whereas in the right panel the scale length is scaled by $f_\text{G} = 2$ compared to the fiducial simulation. The first row displays simulations with a gaseous disc mass halved with respect to the second row ($f_\text{M} = 1/2$), while the third row shows simulations with gaseous disc mass doubled relative to the reference value ($f_\text{M} =2$). Brighter colours represent higher surface density as indicated in the colorbar, which is kept the same in all panels to facilitate the comparison. The panels have the same physical size, $70\times70$ kpc for the face-on projections and $70\times15$ kpc for the edge-on projections. For subsequent analysis (see Sect.~\ref{vertical} and \ref{radial}) it is useful to introduce a characteristic radius $R_d$ representing the size of the gaseous disc. We define $R_d$ as the radius at which the gas surface density drops below 1 M$_{\odot}$ pc$^{-2}$ \citep[see also][]{diteodoro2021}. It is important to note that this definition generally results in $R_d$ differing from the disc scale length $h_\text{g}$ set in the initial conditions. This characteristic radius is shown as a white dashed circle in Figure \ref{density_proj} for the \mwrlittle, \mw \ and \mwrbig \ simulations (second row). The adopted $R_d$ definition captures well the disc extent in all the simulations analysed here.

All nine simulations produce stable discs, with no evidence of bar formation. The resulting disc morphology is similar in all runs despite the different scale lengths, which however influence the overall disc extent. The gaseous disc scale heights range from 0.3 to 0.8 kpc across all simulations and remains approximately constant over 2 Gyr. This scale height increases with radius due to the decreasing gravitational acceleration in the outer regions of the disc, leading to disc flaring. Looking at the face-on projections (top panels) we can appreciate the formation of a multiphase ISM distributed in low density cavities and high density filaments, structures observed in recent JWST observations \citep{watkins2023, thilker2023}. Low-density cavities contain hot gas ($T\approx10^6$ K) and are generated by SNe explosions in the galactic discs. These explosions heat and push the gas towards the edge of the cavities increasing, there, the cold gas surface density. Dense cold clumps are the environment in which new star formation is triggered. The newly formed stars can again inject momentum and energy in the ISM through feedback processes, disrupting cold gas clouds, eventually forming new cavities. These self-regulating mechanisms, implemented in \smuggle, can maintain the SFR stable over several Gyr.

Clustered SN explosions can generate large cavities (that can reach sizes up to 3 kpc, \citealt{egorov2017}) in the ISM, also known as superbubbles \citep[e.g.][]{reyes2014}. These superbubbles may reach the vertical edge of the cold gaseous disc, eventually breaking through and pushing gas from the star-forming disc into the CGM. Stellar feedback processes in \smuggle\ are able to eject gas from the galactic disc, as clearly shown in the edge-on projections of Figure \ref{density_proj} (bottom panels). The majority of these outflows appears to be confined in the first $1-2$ kpc above the disc, but some of the ejected gas can reach distances up to $10-15$ kpc. This ejected gas is a manifestation of the so-called galactic fountain \citep{shapiro1976,fraternali2008}, as typically it has a velocity that is insufficient to escape the galactic potential. As a result, such gas is bound to the galaxy and will eventually fall back onto the disc. Fountain flows are present in all nine simulations as their generation is a natural consequence of the formation of new stars and the associated feedback. The distribution and the intensity of these galactic fountains is analysed in the following sections.

To demonstrate that \smuggle\ is able to regulate the SFR in our simulated galaxies, Figure \ref{sfr} shows the evolution of this quantity as a function of time for the entire simulation set. The early evolution of the SFR should be treated with caution. While we assume the corona is in equilibrium, cooling processes disrupt this balance, causing pressure loss and rapid accretion onto the disc. After about $0.3$ Gyr, the SFR stabilizes due to the self-regulation between star formation and feedback in the \smuggle \ model. The corona also becomes more stable, though it continues to accrete onto the disc through radiative cooling. All the runs display, after this transient period in the first $\sim 300$ Myr (not shown in the figure), an approximately constant SFR spanning the range $2-10\,\,{\rm M_\odot\,yr^{-1}}$. The average level of SFR depends on the initial properties of the disc and SFR variations for each simulation are confined within a factor of $\sim 2$ from this average value. The \mw \ simulation starts with a higher SFR that quickly stabilizes at around $4-5$ M$_{\odot}$ yr$^{-1}$. More specifically, the \mwrlittle \ simulation starts with a higher SFR because of its higher gas surface density (it has the same gas mass contained in a smaller disc) but ends up with a lower SFR compared to the fiducial \mw \ simulation. \mwrbig \ has the opposite behaviour, the SFR grows up over time surpassing the \mwrlittle \ simulation at $t \approx 1\, {\rm Gyr}$. Simulations with a larger disc mass have in general a higher SFR. Looking at the simulations with smaller disc mass: \mwmlittlerlittle \ simulation has a SFR lower than the $f_M=1$ counterpart \mwrlittle, this does not happen for \mwmlittle \ and \mwmlittlerbig. In this case, the SFR is lower in the first 0.8 Gyr and becomes slightly larger for the rest of the simulation.

\begin{figure*}
\centering
\includegraphics[width=\textwidth]{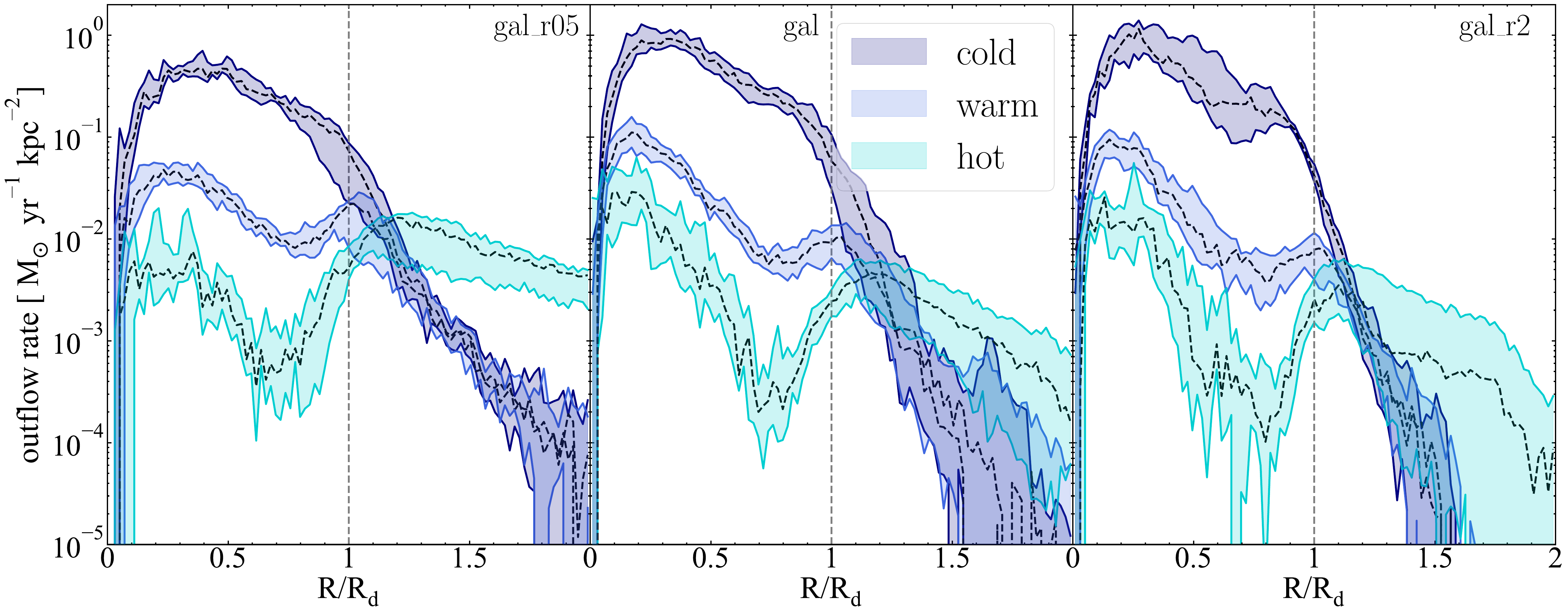}

\caption{Outflow rate per unit area as a function of the cylindrical radius R (in units of the disc size $R_d$) for \mwrlittle \ (left panel), \mw \ (central panel) and \mwrbig \ (right panel) simulations. Each shaded area indicates the outflow rate profiles between the $25^{\text{th}}$ and the $75^{\text{th}}$ percentile, whereas the black dashed lines represent the median values. The rates are divided in cold ($T<10^4$ K, dark blue shaded area), warm ($10^4<T<5\times10^5$ K, blue shaded area) and hot ($T>5\times10^5$ K, light blue shaded area) gas and computed at a height $|z_0|=0.5$ kpc inside a slab over the plane of the disc with a thickness $\Delta z = 0.1$ kpc and with $|v_z|>30$ km s$^{-1}$.}\label{outflow}
\end{figure*}

To better highlight the presence of a galactic fountain cycle in our simulations, Figure \ref{metal_edge} shows edge-on density-weighted projections of the metallicity (left panel), radial velocity (middle panel) and temperature (right panel) of the gas for the fiducial run \mw \ after 0.5 Gyr of evolution. Each panel has a physical size of $70\times70$ kpc. The colour mapping encodes metallicities between $0.2$ Z$_{\odot}$ and $2$ Z$_{\odot}$, radial velocities between $-150$ km s$^{-1}$ and $150$ km s$^{-1}$ and temperatures between $10^2$ K and $10^7$ K (see the colourbar in each panel). As expected, the central region of the disc has a high metallicity ($1-2$ Z$_{\odot}$), which is consistent with the fact that the disc region is where the majority of the star formation is concentrated and stellar evolution can spread the bulk of metals in the ISM. The gas ejected by the galactic fountain has a metallicity consistent with the star-forming disc ($\approx 1$ Z$_{\odot}$), as it is expelled from that region by the action of stellar feedback. The CGM gas has an initial metallicity of $Z= 0.1$ Z$_{\odot}$, but we clearly notice the presence of metal-enriched gas ($\approx0.5$ Z$_{\odot}$) at $\approx 30$ kpc from the disc. This is the result of the mixing of disc gas, which is brought to these distances by the galactic fountain cycle, with the pristine coronal material. The middle panel shows the spherical radial velocity, red-coloured gas is moving radially away from the centre of the galaxy while blue-coloured gas is moving towards it. The galactic fountain cycle is clearly visible: gas that was pushed outside the galaxy (positive radial velocity) is then later re-accreted onto the disc (negative radial velocity). Looking at the metallicity projection we can notice that the gas moving towards the galaxy generally features a lower metallicity than the outflowing gas, which is consistent with the mixing scenario described above. Another important difference between the gas from the disc and the CGM is its temperature (right panel). The ISM is for the majority cold ($T\lesssim10^4$ K) and this also the typical temperature of outflowing gas. As for the metallicity, when the fountain gas travels in the CGM it mixes and accretes material from it, forming an intermediate gas phase that subsequently cools and accretes onto the star-forming disc. As suggested in previous studies \citep[see, e.g.,][]{marinacci2010,hobbs2020, barbani2023}, this fountain-driven gas accretion may represent an important source of fresh gas for star-forming galaxies, directly impacting their evolution. 

The presence of the galactic fountain cycle in each simulation is highlighted in Figure \ref{proj_radial} which shows density-weighted gas spherical radial velocity projections for the 9 simulations (as indicated in each panel). The colour mapping is the same as central panel in Figure \ref{metal_edge}. Each successive row (from top to bottom) increases the mass of the gaseous disc by a factor of 2, while each column from left to right represents a doubling of the disc scale length. The Figure emphasizes that simulations with a higher SFR ($f_M$ = 2 simulations) have stronger ejections of gas from the disc, this results in clouds of cold gas that are pushed at larger distances from the disc and falling back into it after a larger time. 

Figure \ref{high_metal_map} presents a density-weighted projection of metallicity for the \mw \ simulation at $t=0.8$ Gyr, considering only high-metallicity gas ($Z>0.9 \,Z_{\odot}$), with the same colour mapping of the central panel of Figure \ref{metal_edge}. The inflowing gas (blue-coloured) is enriched in metals, highlighting that some fraction of it originates from the disc. Therefore, a significant fraction of the inflowing gas stems from the galactic fountain process and is on its journey back onto the disc. This reinforces the idea that metal-enriched gas is recycled within the system, as there are no external sources of metallicity in the simulation.

\subsection{Vertical galactic fountain distribution}\label{vertical}

Galactic fountains play a fundamental role in galaxy evolution, by interacting with the CGM and bringing new gas to the disc \citep[e.g.][]{marasco2022, li2023}. In this Section we want to analyse how gaseous outflows and inflows are distributed over the galactic disc. Outflow and inflow rates are computed inside a slab centred at a height $z_0$ and with a thickness $\Delta z$ above and below the plane of the disc (see also \citealt{barbani2023}). If a gas cell inside the slab meets all the required conditions (see below), it is counted as inflowing/outflowing gas and its contribution to the total mass rate is summed up as 
\begin{equation}
\dot{M}_{\text{in/out}}=\sum_i \frac{m_i v_{z,i}}{\Delta z},
\end{equation}\label{infoutf}
where $m_i$ is the mass of the gas cell and $v_{z,i}$ is the $z-$component of its velocity.
We set a velocity threshold $v_{\text{thres}}=30$ km s$^{-1}$, this value is chosen as an optimal compromise, ensuring that turbulent motions (tens of km s$^{-1}$) are not misclassified as inflows/outflows, while still accounting for most outflows with velocities in the range 50 km s$^{-1}$<$v_z$<100 km s$^{-1}$. If $v_{z,i} z_i > 0$ and $|v_{z,i}| > v_{\text{thres}}$, where $z_i$ is the vertical distance from the disc plane of the gas cell, the gas in the cell is considered as a part of an outflow, whereas if $v_{z,i} z_i < 0$ and $|v_{z,i}| > v_{\text{thres}}$ it is counted as inflowing gas.

\begin{figure*}
\centering
\includegraphics[width=\textwidth]{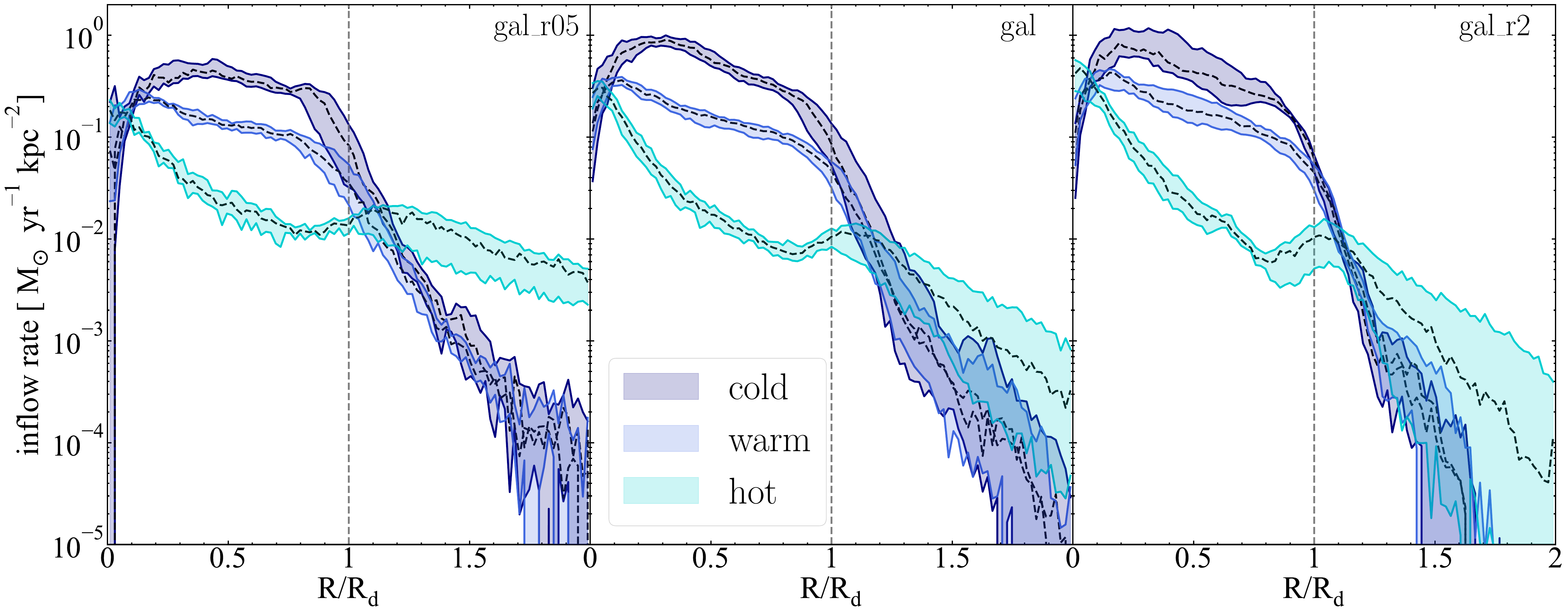}

\caption{The same as Figure \ref{outflow}, but for inflow rates.}\label{inflow}
\end{figure*}

Figure \ref{outflow} shows the outflow rate per unit area as a function of cylindrical radius normalized by $R_d$ for the \mwrlittle \ (left panel), \mw \ (middle panel) and \mwrbig \ (right panel) simulations, computed at $|z_0|=0.5$ kpc. This height corresponds to $\sim 1-2$ gaseous disc scale height and ensures that we focus on the region near the mid-plane, where inflows and outflows interact most directly with the star-forming disc. The disc size $R_d$ is defined as the radius at which the gas surface density drops below 1 M$_{\odot}$ pc$^{-2}$ (see Figure \ref{density_proj}), resulting in $R_d$ values of $\approx 20$ kpc for the fiducial galaxy, $\approx 15$ kpc for \mwrlittle \ and $\approx 27$ kpc for \mw. These disc sizes are used to normalize the $x$-axis, which extends in the range $0 \leq R/R_d \leq 2$. The gas has been divided in cold ($T<10^4$ K, dark blue shaded area), warm ($10^4<T<5\times10^5$ K, blue shaded area) and hot ($T>5\times10^5$ K, light blue shaded area). The outflow rate profiles have been computed at each time in the simulation with a temporal resolution $\Delta t=10$ Myr (temporal difference between each snapshot of the simulation), the radial bins width is $\Delta R=0.02$ in $R/R_d$ units. For each panel we computed the 25$^{\text{th}}$ and 75$^{\text{th}}$ percentiles of the resulting radial profiles and displayed their extent in the plot as the shaded coloured area, whereas the dashed black line indicates the median value. We notice that the outflows are dominated by the cold phase, which is on average at least one order of magnitude larger than the warm phase and two orders of magnitude larger than the hot one. Thus, the cold phase dominates the mass budget in the gas swept by SNe. The outflow rates are decreasing with the radius, following the distribution of gas and stars. At $R\lesssim 0.4 R_d$, where more star formation is concentrated, the outflows are stronger. As expected, the distribution of the outflows is connected to the extension of the disc. All the simulations show a similar shape distribution with a break point at the edge of the disc $R_d$, beyond this radius the outflow rates drop significantly. In the fiducial simulation, the outflow rate is higher than $10^{-1}$ M$_{\odot}$ yr$^{-1}$ kpc$^{-2}$ until $R_d\approx20$ kpc at all times. In \mwrlittle, that has a halved disc scale length, the outflows arrive to $\approx 15$ kpc and in \mwrbig \ they extend up to $\approx30$ kpc. At $t=1$ Gyr these three runs have a very similar SFR ($4-5$ M$_{\odot}$ yr$^{-1}$, see Figure \ref{sfr}) and the intensity of the outflow rate in the first kpc is similar, but in \mwrbig \ the outflow are more extended. Therefore, the global outflow rate is higher in this system, with direct consequences on its evolution. For instance, more outflows can interact with the hot gas in the CGM, consequently increasing also the accretion rate. During the 2 Gyr evolution there is not a large change in the outflow rate profile and therefore the shaded areas are very close to the median value.

\begin{figure*}
\centering
\includegraphics[width=\textwidth]{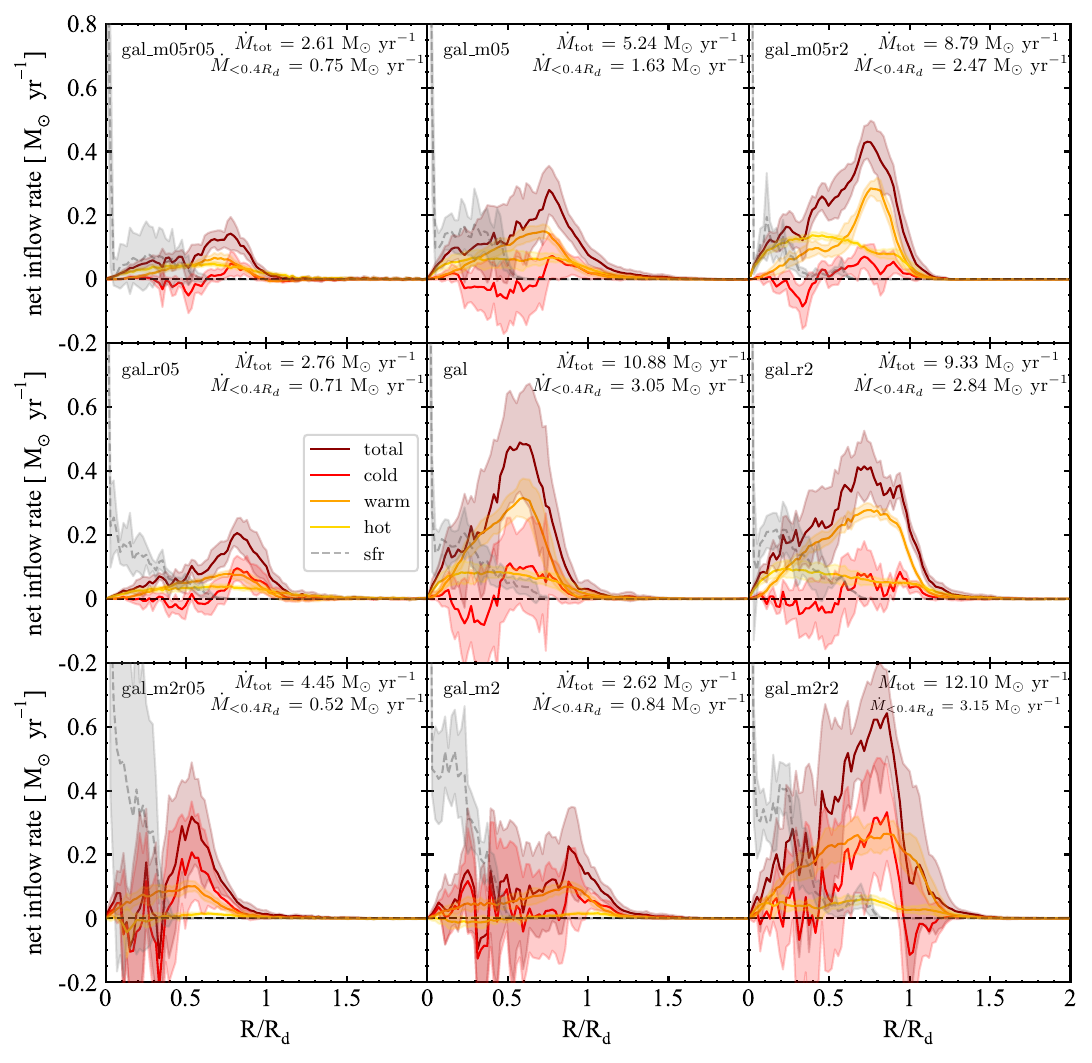}

\caption{Net vertical inflow rate, computed as $\dot{M}_{\text{net}}=\dot{M}_{\text{in}}-\dot{M}_{\text{out}}$, as a function of cylindrical radius $R$ (in units of $R_d$) for the 9 simulations divided in cold ($T<10^4$ K, red line), warm ($10^4<T<5\times10^5$ K, orange line) and hot ($T>5\times10^5$ K, yellow line) gas. The dashed grey line is the SFR profile. The solid lines represent the profile averaged over the entire simulation and the shaded areas are the standard deviations. The rates are evaluated inside a slab over the plane of the disc with a width $\Delta z = 0.1$ kpc  at a height $|z_0|=0.5$ kpc. Each point represents the gas net inflow rate within a circular annulus with a radial width of 0.02 in $R/R_d$ units and considering only gas cells with $|v_z|> v_{th} = 30$ km s$^{-1}$. In each panel we have indicated the average net vertical rate at $R<0.4R_d$, this is roughly the region of the disc where $90\%$ of the SFR is concentrated in all the 9 simulations. The net inflows can extend beyond $R_d$ supplying the outer disc with fresh gas for star formation.}\label{netrate}
\end{figure*}

\begin{figure*}
\centering
\includegraphics[width=\textwidth]{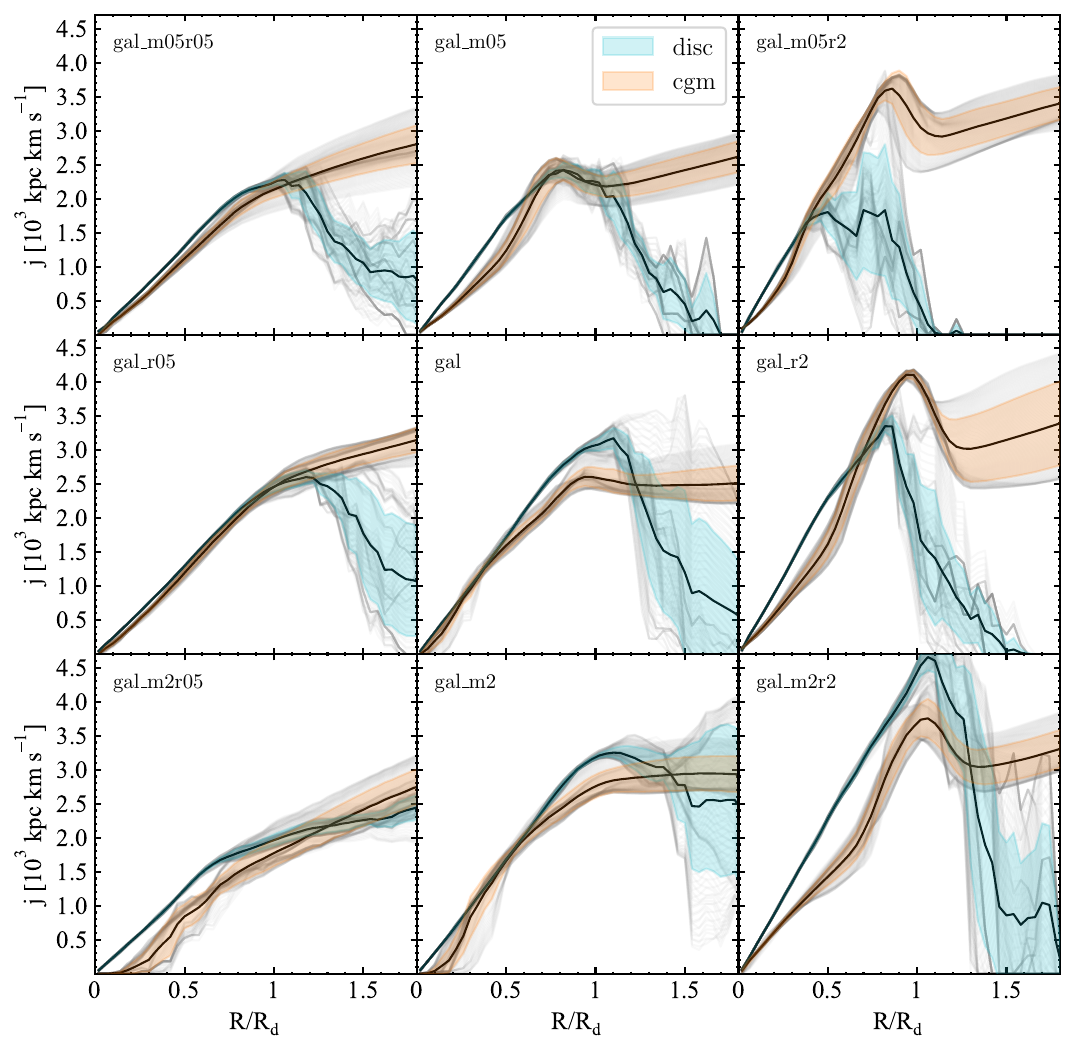}

\caption{Specific angular momentum, computed as \( j = J / M \), where \( J \) is the total angular momentum and \( M \) is the total mass of each respective component, for the gaseous disc and the CGM computed as a function of cylindrical radius $R$ (in units of $R_d$) for the 9 simulations. The angular momentum is measured near the disc (\( |z| < 1 \) kpc), with CGM gas selected by \( Z < Z_{\odot} \) and disc gas by \( Z \geq Z_{\odot} \). In each panel, the grey lines represent the specific angular momentum profile at a given time averaged in a time span of 100 Myr. The black line is the average specific angular momentum and the blue (red) shaded area represents the standard deviation of the disc (CGM) specific angular momentum distribution. Each panel represents a different combination of disc mass and scale length, showing how these parameters influence the angular momentum distribution. Across all models, the disc maintains a systematically higher specific angular momentum than the CGM in the inner regions following the rotation curve, while at \( R/R_d \gtrsim 1 \), the disparity between the two components becomes more variable, marking the transition where the disc fades.}\label{angmom}
\end{figure*}

\begin{figure*}
\centering
\includegraphics[width=\textwidth]{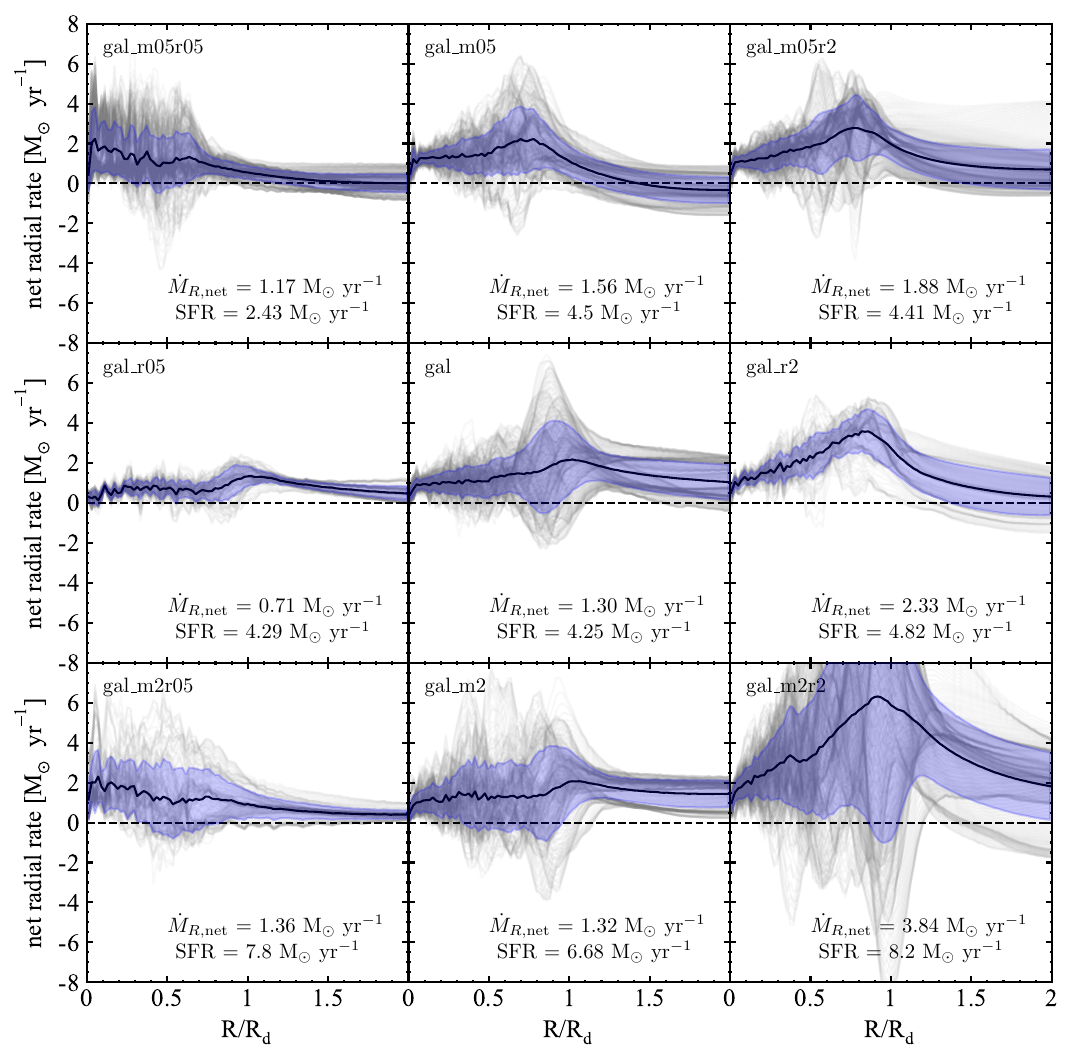}
\caption{Gas net radial mass rate for the simulations analysed in this work. In each panel, the grey lines represent the mass rate profile at a given time averaged in a time span of 100 Myr. The black line is the average mass rate and the blue shaded area represents the standard deviation. On the bottom of each panel the net radial mass rate averaged within the radial range  $0 \leq R/R_d \leq 1$ and the average SFR, computed in the temporal range $0.3<t<2$ Gyr, are indicated.}
\label{radial_mass}
\end{figure*}

\begin{figure*}
\centering
\includegraphics[width=\textwidth]{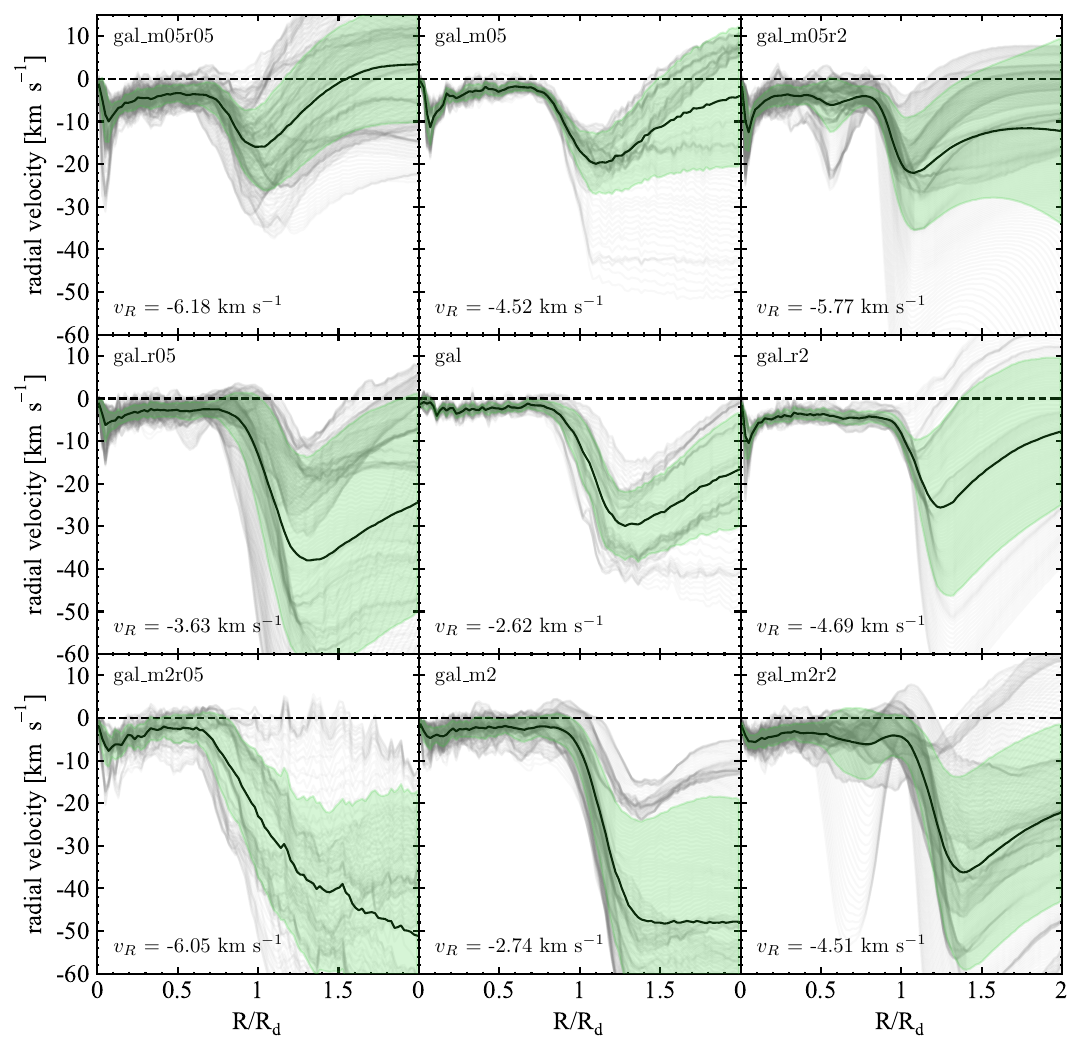}
\caption{Gas cylindrical radial velocity for the simulations analysed in this work. In each panel, the grey lines represent the (mass weighted) radial velocity profile at a given time averaged in a time span of 100 Myr. The black line is the average cylindrical radial velocity and the green shaded area represents the standard deviation. On the bottom left corner the radial velocity averaged within the radial range  $0 \leq R/R_d \leq 1$ is indicated.}
\label{radial_vel}
\end{figure*}

Figure \ref{inflow} shows inflow rate per unit area as a function of cylindrical radius normalized by $R_d$ for the same simulations and in the same temperature range of Figure \ref{outflow}. As this quantity is being computed at a relatively close distance to the disc mid-plane, i.e. $|z_0|=0.5$ kpc, the accreted gas has time to cool down before entering the gaseous disc, therefore the inflowing gas is mostly cold. Similarly to the outflow rates, the extension of the inflow rate distribution is connected to the size of the disc, more extended discs have more extended inflows, showing a break in the distribution at $\approx R_d$, after which the inflow rates decline rapidly. This happens because a large fraction of the inflows is gas that was beforehand ejected from the disc and subsequently falls back into it. The analogy between outflow and inflow rate profiles is again suggesting an essential role of the galactic fountains in setting the spatial distribution of the gas onto the disc. In the \mw \ simulation the inflow rate arrives to a lower radius with respect to \mwrbig, indicating that the majority of the accreted gas comes from the recycling of the ejected gas in a galactic fountain cycle. The other 6 simulations (not shown) behave in a similar way.

Our goal is to examine the net vertical mass inflow of gas onto the disc, considering that a significant portion of this inflowing gas was previously ejected from the disc as part of the galactic fountain process. Figure \ref{netrate} shows the average net inflow rate, computed using Eq. (\ref{infoutf}) as $\dot{M}_{\text{net}} = \dot{M}_{\text{in}} - \dot{M}_{\text{out}}$, divided in cold (red line), warm (orange line) and hot (yellow line) gas for the 9 simulations analysed in this work (as indicated in each panel). The rates are computed at a height of $|z_0|=0.5$ kpc from the disc plane with a $\Delta z = 0.1$ kpc width. Together with the net inflow rate we show the SFR radial profile (dashed grey line). As in the previous figure, the disc is divided into radial bins of width 0.02 in $R/R_d$ units.

Figure \ref{netrate} emphasizes the connection between accretion from the CGM and galactic disc structure. In all the simulations, the net inflow rates exhibit a bell-shaped distribution. These accretion rates extend up to $\approx 1.2\,R_d$ in all the simulations, which corresponds to $\approx 24$ kpc in the $f_G=1$ simulations, $\approx18$ kpc in the $f_G=0.5$ simulations and $\approx32.5$ kpc in the $f_G=2$ simulations, following the respective discs sizes $R_d$. It is worth to notice that the peak of the total net inflow rate is not located at the centre of the galaxy, but is instead skewed towards larger radii. The peak distance from the centre is increasing with disc size and is typically around $0.75\,R_d$. The grey dashed line shows the SFR as a function of radius, which has a central peak and also a bell shape but is more concentrated in the inner region of the disc, where the gas surface density is higher. If the gas were to fall back to roughly the same radial position from which it was ejected, the inflow distribution would closely resemble the SFR distribution. Instead, the majority of the gas is accreted in the outer regions of the disc. This happens because the orbital paths of galactic fountain clouds increase with radius, due to the progressively  weaker gravitational attraction moving towards the external regions of the galaxy \citep[e.g.][]{li2023}. Additionally, the region of the CGM affected by the fountain clouds passage -- and where the warm gas phase is generated -- expands with their orbital path, facilitating an easier accretion into the disc. The total net accretion rate $\dot{M}_{\text{tot}}$ ($R<R_d$) is indicated in each panel. This total accreted gas would be enough to sustain the SFR (Figure \ref{sfr}), a fraction of this accretion in the external regions is used to grow the disc inside-out (see Section \ref{insideout}). While most of the gas is accreted in the outskirts of the disc, a fraction of it falls in the central region ($R<0.4\,R_d$), in each panel of Figure \ref{netrate} we indicated the average value of the net rate in this region. $90\%$ of the SFR is located at $R<0.4\,R_d$, thus gas accreted in these central regions can be directly available for the formation of new stars, whereas gas accreted at $R>0.4\,R_d$ cannot be immediately used for star formation as the average gas density is not high enough. This average net rate, while not negligible, is still not sufficient to entirely sustain the SFR of the galaxy (see Figures \ref{sfr} and \ref{radial_mass}). Therefore there needs to be an alternative mechanism that feeds the star formation in the centre. The accretion budget and the comparison with the SFR are discussed in Section \ref{radial}.

When examining the net inflow rate separated in hot, warm and cold components, we can notice several interesting features. We have seen that the outflow rate of warm and hot gas is almost negligible (Figure \ref{outflow}), indicating that the net warm/hot inflow primarily originates from the CGM rather than just being gas previously ejected from the disc. In fact, as shown in \citet{barbani2023}, the warm gas phase is generated by the passage of the cold fountain clouds through the hot CGM. In all simulations, the hot gas has an almost flat distribution that slightly decreases with increasing radius, with values ranging from $0$ to $\sim0.15$ M$_{\odot}$ yr$^{-1}$. This likely represents gas directly accreted from the hot CGM via a cooling flow, showing no clear correlation with the cold and warm gas distributions. The warm and cold gas, however, are more closely connected, with their net inflow peaks occurring at similar radii. The net cold inflow rate shows more discontinuous peaks and it is not consistently positive\footnote{In the Figures showing net mass rates, we use the convention that a positive net rate represents gas accretion onto the disc, whereas a negative rate is interpreted as gas ejection from the disc.}. As shown in Figure \ref{outflow}, nearly all of the ejected gas is in the cold phase, leading to regions of the disc where the outflow rate exceeds the inflow rate (resulting in a negative net rate) and vice versa. This does not happen for the net warm inflow rate, which remains consistently positive and generally higher than the net cold inflow rate. The spatial distribution of cold and warm inflows appears interconnected, suggesting that cold outflows may enhance the formation of inflowing warm gas. As discussed extensively in \citet{barbani2023}, mixing between the cold fountain clouds and the hot CGM can reduce the cooling time of the gas mixture, allowing the gas to cool more rapidly and accrete onto the disc, thereby supplying fresh gas to sustain star formation.

The different disc sizes and gaseous masses also result in variations in the net inflow distribution. The simulations with the smaller discs (\mwrlittle, \mwmlittlerlittle, \mwmbigrlittle) show the lowest net inflow rates, this is a combination of both a lower SFR (see average SFR indicated in each panel of Figure \ref{radial_mass}) and a smaller extent of the disc. The hot gas accretion rate increases with disc size, reflecting the slow accretion of the galactic corona; a larger disc is more capable of capturing and accreting more hot gas. The cold gas rate, that derives from stellar feedback, is also lower in smaller discs due to their lower SFR, which in turn reduces the generation of warm gas, formed from the mixing and condensation of CGM into the galactic fountain cloud. Consequently, the rates increase for larger discs, which also present on average a higher SFR. Therefore, if the mass of the gaseous disc decreases, gas accretion diminishes for the same reasons.

\subsection{Radial gas motions}\label{radial}

Figure \ref{netrate} illustrates that the majority of the gas is accreted at a significant distance from the galactic centre ($\approx 0.75 R_d$), near the edge of the gaseous disc. Star forming galaxies need to acquire gas from the external environment to keep forming stars with an almost constant SFR \citep[e.g.][]{cignoni2006, isern2019,mor2019}, as the gas within the disc is not sufficient. The majority of stars form in the central regions of the disc, where gas density is the highest. Therefore, the accreted gas must travel from the disc outskirts to the centre in order to be accessible for the formation of new stars \citep{schmidt2016,diteodoro2021}.

\begin{table*}
	\centering
	\caption{Summary of the SFR and the net gas accretion rate $\dot{M}_{\text{SFR}}=\dot{M}_{z, \text{net}}(R<0.4R_d) + \dot{M}_{R, \text{net}}$ for all simulations analysed in this work.}\label{sfr_param}
	\begin{tabular}[t]{cccccccccc}
	
		\hline
        \hline
		Simulation name & \mwrlittle & \mw & \mwrbig & \mwmlittlerlittle & \mwmlittle & \mwmlittlerbig & \mwmbigrlittle & \mwmbig & \mwmbigrbig\\
        \hline
	    \rule{0pt}{3ex}  
		$\dot{M}_{\text{SFR}}$ [M$_{\odot}$ yr$^{-1}$] & 1.42 & 4.35 & 5.2 & 1.92 & 3.19 & 4.35 & 1.9 & 2.16 & 7 \\
        \hline
        \rule{0pt}{3ex}  
		$\text{SFR}$ [M$_{\odot}$ yr$^{-1}$] & 4.29 & 4.25 & 4.82 & 2.43 & 4.5 & 4.4 & 7.8 & 6.68 & 8.2 \\
		\hline
        \hline
	\end{tabular}
    \tablefoot{The comparison highlights the relationship between gas replenishment from the CGM and the sustenance of star formation over the 2 Gyr time span probed by the simulations analysed in this work.}
\end{table*}

A key aspect of understanding radial gas motion within the disc is the distribution of angular momentum in both the gaseous disc and the CGM. Figure \ref{angmom} presents the specific angular momentum profiles for the nine different galaxy models, computed as \( j = J / M \), where \( J \) is the total angular momentum and \( M \) is the total mass of each respective component. The angular momentum is evaluated near the disc (\( |z|< 1 \) kpc), with CGM gas selected by metallicity (\( Z < Z_{\odot} \)) and disc gas by (\( Z \geq Z_{\odot} \)). The figure reveals a consistent trend: in the inner regions the specific angular momentum of the gaseous disc is systematically higher than that of the CGM throughout the entire time span of the simulations. Within \( R_d = 1 \), the disc angular momentum increases linearly closely following the trend expected given the disc rotation curve, indicating full rotational support. Beyond this point, the disc component fades, and its angular momentum declines. A similar trend is observed in the inner CGM, though at lower values, suggesting some degree of co-rotation with the gaseous disc and a spin up of the corona in the regions close to the disc, likely driven by galactic fountains, as also indicated by cosmological simulations \citep[e.g.,][]{grand2019}. At larger radii (\( R/R_d \gtrsim 1 \)), the disc angular momentum diminishes, while the CGM angular momentum initially decreases before slowly increasing outwards, marking the transition to a regime where the coronal gas ceases to co-rotate with the disc. The impact of galaxy properties is evident in the figure. In the \( f_g = 2 \) simulations, increasing the total gas mass of the disc leads to a noticeable increase in the disc specific angular momentum at all radii. In the other cases, changes in gas mass do not significantly affect the absolute angular momentum values. Variations in disc scale length also influence the angular momentum distribution, with larger scale lengths generally corresponding to higher disc angular momentum at fixed gas mass. Despite these differences, the higher angular momentum of the disc relative to the CGM in the inner regions remains a robust feature across all models.

To investigate the radial migration of gas we have computed the radial mass rate in each radial bin as $\dot{M}_{R\text{, in/out}}=\sum_i v_{R, i} m_i/ \Delta R$, where $v_{R, i}$ is the cylindrical radial velocity of the $i$-th gas particle and $\Delta R$ is the radial bin size. The particles are selected within a slab centred on the disc plane ($z_0=0$ kpc) with a thickness of $\Delta z=0.5$ kpc to ensure that only disc particles are included. The radial outflow (inflow) rate is determined by considering particles with $v_R>0$ ($v_R<0$). The net mass rate is computed as $\dot{M}_{R,\text{net}}=\dot{M}_{R,\text{in}}-\dot{M}_{R,\text{out}}$. The radial velocity is computed averaging the cylindrical radial component of the velocity weighted by the mass of each gaseous cell. For both quantities, the disc is divided in annuli with a spatial extent $\Delta R = 0.02$ in $R/R_d$ units and the profiles are computed in the radial range $0 \leq R/R_d \leq 2$.

Figure \ref{radial_mass} shows the net radial mass rate of the gas for the 9 simulations, as indicated in each panel. The grey lines represent the net inflow rate profile at individual times, while the black line shows the average profile over a 2 Gyr time span, with the shaded area indicating the standard deviation. The average value of the gas mass rate is positive throughout the entire disc, indicating a significant radial mass flow towards the inner regions. When the accreted material has a specific angular momentum that is different than the one in the disc, radial flows will inevitably arise owing to angular momentum conservation \citep{mayor1981}. Since the gas accreted from the CGM has a rotational velocity lower than that of the disc, it will naturally tend to migrate towards the inner regions of the disc, forming a radial flow that feeds the star formation. This is apparent in Figure \ref{angmom}, which shows that the specific angular momentum of the CGM is systematically lower than that of the disc in the inner regions. This discrepancy drives angular momentum redistribution and facilitates radial transport of gas towards the centre. 

Generally, the gas net mass rate is near zero in the disc outskirts and it moves towards more positive values (indicating inflows) in the inner regions of the disc across all simulations, eventually approaching zero again at the centre. This mass inflow becomes less prominent in the inner regions for two main reasons: (i) the higher gas density at the centre and (ii) more importantly, the fact that a portion of the gas is consumed to form stars in these central regions. A consistent pattern emerges in all the simulations, a peak in the mass rate that is generally found roughly at the edge of the disc ($R\approx R_d$), except for simulations \mwmbigrlittle \ and \mwmlittlerlittle. The cumulative star formation rate of the galaxy follows a similar pattern, exhibiting a decline for radii $R/R_d \lsim 0.4$, where the large majority of the SFR is located. This similarity in trends emphasizes that star formation primarily takes place in the inner disc regions, with the radial mass flux diminishing towards the centre as it feeds star formation. In all simulations, the SFR remains higher than the radial net inflow rate. This discrepancy is compensated by vertical inflows from the CGM, which, together with radial mass transport, sustain the SFR in the disc (see below). When comparing simulations with the same gas disc scales, we observe values of $\dot{M}_{R,\text{net}}$ within a factor of 2, except for the $f_G=1$ simulations where the differences in the average mass rate value are almost negligible. The primary difference arises when the size of the gaseous disc is changed: the gas mass rate increases with larger disc sizes. A larger disc extends into regions where the CGM has lower angular momentum, facilitating easier inward motions towards the centre of the disc. 

Overall, while the simulations show a non-negligible net inflow rate, this rate is generally lower than the average SFR. For instance, the net radial inflow rate is 1.30 M$_{\odot}$ yr$^{-1}$ for the fiducial run and 0.71 M$_{\odot}$ yr$^{-1}$ and 2.33 M$_{\odot}$ yr$^{-1}$ for \mwrlittle \ and \mwrbig \ respectively, compared to the average SFR of 4.5 M$_{\odot}$ yr$^{-1}$, 4.29 M$_{\odot}$ yr$^{-1}$ and 4.5 M$_{\odot}$ yr$^{-1}$ for these cases. Thus, radial mass motion towards the centre alone may not fully sustain the SFR over time. However, it is important to note that not all gas is accreted at the edge of the disc; a significant fraction, approximately $25\%$ of the total accreted gas, is directly deposited in the inner regions ($R<0.4R_d$, see Figure \ref{netrate}), where $90\%$ of star formation takes place. Combined with radially transported gas, this could be sufficient to sustain the global SFR of the galaxy.
Therefore, for the SFR to be fueled by these two phenomena in conjunction we would need to have $\dot{M}_{\text{SFR}}=\dot{M}_{z, \text{net}}(R<0.4R_d) + \dot{M}_{R, \text{net}}\approx$ SFR. In fact, this is true for most of the 9 simulations analysed in this work (see Table \ref{sfr_param}). For instance, in \mwmlittlerlittle \ we find $\dot{M}_{\text{SFR}}\approx 1.92$ M$_{\odot}$ yr$^{-1}$, which has to be compared with the average $\text{SFR}\approx 2.43$ M$_{\odot}$ yr$^{-1}$. Even though strictly speaking the rate of replenishment of gas is below the SFR, the two values are comparable. Moreover, we have to consider that $\dot{M}_{\text{SFR}}$ and SFR are averaged over a 0.3-2 Gyr time span, meaning that our analysis is smoothing what occurs over the evolution of the galaxy. Therefore, these values should be interpreted with some caution, but nonetheless they provide insight into whether the formation of new stars can be sustained by gas accretion from the CGM. Other systems show a similar behaviour. Overall, as seen in Table \ref{sfr_param}, the majority of the simulations have $\dot{M}_{\text{SFR}}\approx\text{SFR}$, however there are some exceptions: in \mwrlittle \ we find  $\dot{M}_{\text{SFR}}\approx 1.42$ M$_{\odot}$ yr$^{-1}$ which accounts for less than half of the average SFR ($\approx 4.29$ M$_{\odot}$ yr$^{-1}$). Similarly, in \mwmbigrlittle \ and \mwmbig, the SFR deviates from $\dot{M}_{\text{SFR}}$, as it is partially sustained by the larger initial gas reservoir in the disc at the start of the simulations.

\begin{figure*}
\centering
\includegraphics[width=\textwidth]{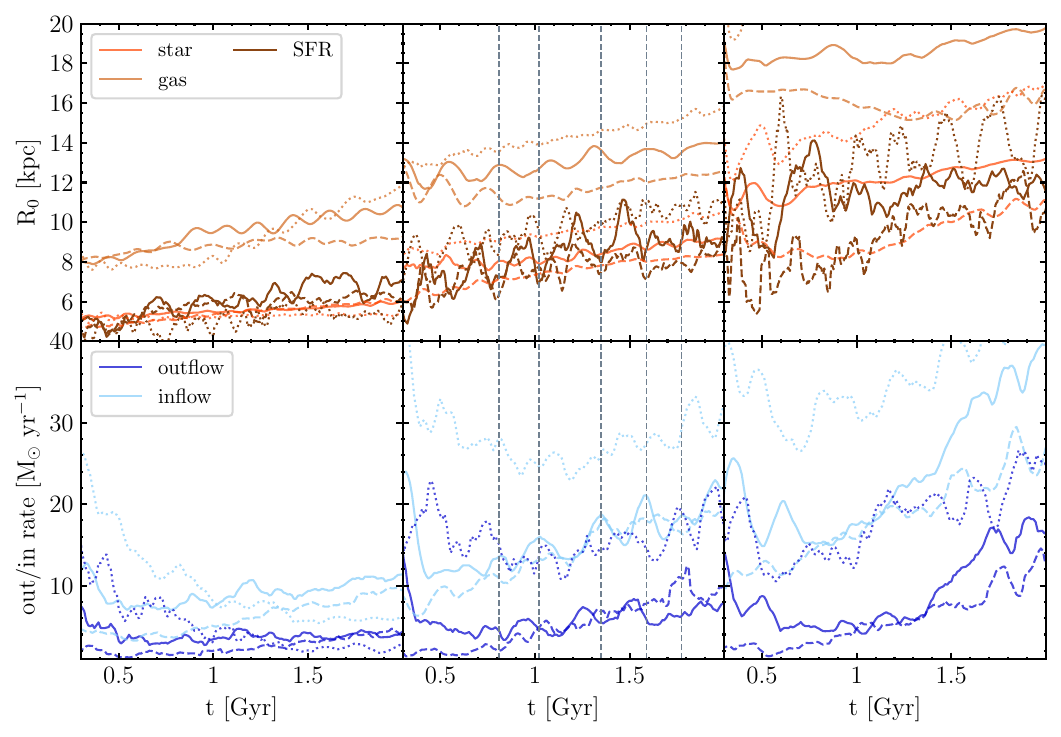}

\caption{Top panels: star (orange line), gas (light brown line) and star formation rate (brown line) scale length for the \mwrlittle \ (solid), \mwmlittlerlittle \ (dashed), \mwmbigrlittle \ (dotted, left panel), \mw \ (solid), \mwmlittle \ (dashed), \mwmbig \ (dotted, central panel) and \mwrbig \ (solid), \mwmlittlerbig \ (dashed), \mwmbigrbig \ (dotted, right panel) simulations as a function of time. These scale lenghts are computed as the radius that encloses $90\%$ of stars/gas mass or SFR.
Bottom panel: total outflow (blue) and inflow (light blue) rates of the same simulations of the top panels as a function of time. The simulations with $f_M=0.5$ are indicated with a dashed line and the simulations with $f_M=2$ with a dotted line. In the central panels the peak in the inflow rate of the \mw \ simulation are highlighted with dashed grey vertical lines. Each of the scale is connected and they have an up-and-down trend linked to the gas outflow/inflow cycle.}\label{gas_scale}

\end{figure*}

Figure \ref{radial_vel} shows the average radial velocity, with values ranging between -4 and -6 km s$^{-1}$. All the simulations exhibit a similar radial velocity profile: the velocity remains almost constant within the gaseous disc ($R<R_d$), rapidly decreases outside the disc and then start to slightly increase again (except in \mwmbigrlittle \ and \mwmbig). This profile shape appears to be independent of the disc structural characteristics. However, the magnitude of the velocity slightly changes between discs of different sizes. Specifically, the radial velocity increases with disc size, similarly to the mass rate. The average velocity inside the disc increases between $f_M=1$ and $f_M=0.5$ simulations, when the mass of the gaseous disc is smaller, probably  because the lower average ISM density allows gas to flow more easily toward the inner regions. The velocity is larger in the outskirts, outside the galactic disc, where gas is accreted radially on the edge of the galaxy, allowing the disc to grow in size over time (see Figure \ref{gas_scale}). Moving inward, the radial velocity decreases and approaches zero at the centre of the disc. 

\begin{figure}
\centering
\includegraphics[width=\columnwidth]{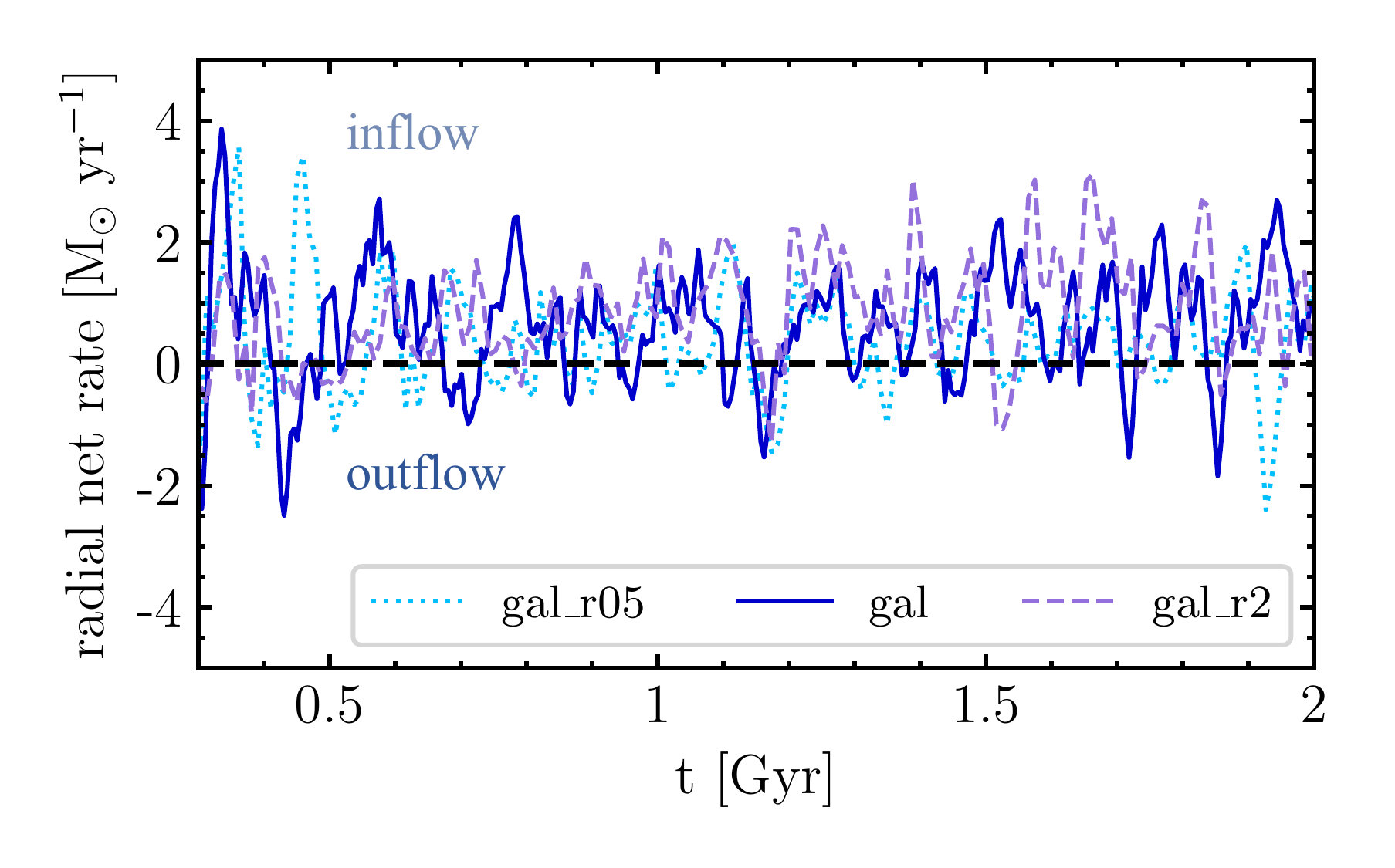}

\caption{Radial gas net rate as a function of time for the \mwrlittle \ (dotted line), \mw \ (solid line) and \mwrbig \ (dashed line) simulations. At each time the radial net rate averaged within the disc, at $R<R_d$ and $|z|<0.5$ kpc, is displayed. Overall the radial net rate is mostly positive, indicating a flow of gas towards the centre of the disc.}\label{radial_net_rate}

\end{figure}

Figure \ref{radial_net_rate} presents the radial net rate of the gas as a function of time for the three simulations with the standard gaseous disc mass (\mwrlittle, \mw \ and \mwrbig), the other 6 simulations are not shown as they all have a very similar trend. For each simulation, we computed the total radial net rate by averaging it over the whole galactic disc (i.e. at $R<R_d$ and $|z|<0.5$ kpc) at each time (with a temporal resolution $\approx 10$ Myr). The trend is consistent across all simulations, alternating between periods of negligible or negative net rate and periods of positive net rate. This fluctuation is likely connected with the stellar feedback cycle; when SNe explode the net rate is dominated by the outflowing gas. As negative stellar feedback reduces star formation, thereby decreasing the injection of energy from new SNe, fresh gas can flow towards the centre once again, fueling star formation and repeating the cycle. Overall, the gas predominantly flows towards the centre (positive net rate), suggesting that the radial gas flow phenomenon is independent of the gaseous disc size, given that the rotational velocity of the coronal gas has the same values at the beginning of all simulations.

To sum up, the simulations reveal a general inflowing trend that, while insufficient to sustain the star formation in the centre alone, is nonetheless crucial for feeding the inner regions of the galaxy, working in conjunction with vertical gas accretion in the central regions ($R<0.4R_d$).

\subsection{Inside-out evolution}\label{insideout}

The vertical accretion in the outer regions of the disc, coupled with the subsequent radial motions, drives the gas from the periphery towards the disc centre. Inevitably, this accretion increases the gas density in progressively more outer regions, consequently enhancing star formation activity in these external areas. Over time, this mechanism enables the disc to grow from the inside to the outside, creating what is known as inside-out evolution \citep[e.g.][]{firmani2009, brooks2011, pezzulli2016, li2023}. To explore the relationship between gas accretion and disc growth, we analysed the evolution of different scale lengths over time and their connection to the total outflow/inflow rates. The top panels of Figure \ref{gas_scale} show the evolution of the scale lengths of stars (orange line), gas (light brown line) and SFR (brown line) for the 9 simulations as a function of time. These scale lengths are determined by computing the radius which encloses $90\%$ of the stars/gas mass and of the SFR, with each value smoothed using a moving average over a 100 Myr interval. From left to right each panel shows the $f_G=0.5$, $f_G=1$ and $f_G=2$ simulations respectively, with solid lines indicating the $f_M=1$ runs, dashed lines indicating the $f_M=0.5$ runs and dotted lines indicating the $f_M=2$ runs. All the simulations show a periodic trend with oscillations that can be linked to the typical time scale of a fountain cycle in Milky Way-like galaxies ($t_{\text{fount}}\approx 100$ Myr, \citealp[e.g.][]{fraternali2014}). Over a span of 2 Gyr all three scale lengths have increased. 

We start by focusing on the fiducial run \mw \ (solid lines, central panel): the gas scale length has expanded from $\approx 12$ kpc to $\approx 14$ kpc, the star scale length from $\approx8$ kpc to $\approx9$ kpc and the SFR scale length from $\approx6$ kpc to $\approx 9$ kpc. The gas scale length remains consistently larger than the stellar scale length by about a factor of $\approx 2$. Notably, both the gas and stellar discs exhibit a concurrent growth, expanding steadily over a 2 Gyr period. The stellar scale shows higher peak frequencies, with approximately two peaks for every peak in the gas scale length. Interestingly, the peaks in the SFR do not coincide with those of the gas and star scales; rather, they tend to align with depressions in the gas scale length. The SFR exhibits the greatest amplitude variation, whereas the stellar scale remains more stable, with peaks that nearly disappear in the other simulations. An increase in the SFR spatial extent implies that more stars are being formed in the outer regions of the disc compared to previous times, leading to more outflows ejected from these areas. Consequently, these episodes of gas ejection may reduce the gas distribution scale length, with a fraction of these outflows eventually falling back onto the disc, thus increasing the gas scale and repeating the cycle again. The \mwmlittle\ (central panel, dashed lines) and \mwmbig\ (central panel, dotted lines) simulations, with a gaseous disc mass halved and doubled, respectively, have similar behaviours. The values of the three scales are similar to \mw, with the \mwmlittle \ scales slightly lower and the \mwmbig \ scales slightly larger than the fiducial run scales. Also in these cases the peaks in the gas scale and in the SFR scale are misaligned. Similar features can be seen in the other simulations. The simulations with $f_G=0.5$ (left panel) show lower values in all the three scales as expected and they are all increasing over the 2 Gyr of the simulations. In the $f_G=2$ runs (right panel) the scales are higher than in the $f_G=1$ runs, \mwrbig \ gas scale is passing from $\approx 18$ kpc to $\approx 20$ kpc, whereas in \mwmlittlerbig \ the gas scale stays almost constant at $16$ kpc. Overall, the increasing trend in these scale lengths in all the simulations over time indicates the growth in the size of the star-forming disc, which is consistent with an inside-out evolution of the disc.

In the bottom panels, we display the total outflow (dark blue line) and inflow (light blue line) rates for the 9 simulations, computed at $z_0=0.5$ kpc, $\Delta z = 0.1$ kpc and $R<R_d$. In general, the inflow rate is larger than the outflow rate. In the fiducial run \mw \ (central panel, solid line) the net inflow rate is $\dot{M}_{\text{net}}=\dot{M}_{\text{in}}-\dot{M}_{\text{out}}\approx10$ M$_{\odot}$ yr$^{-1}$. This net inflow is larger than the average SFR observed in the fiducial simulation (see Figure \ref{sfr}), confirming that the formation of new stars is largely sustained by gas accretion from the CGM, the same applies to the other simulations. An important aspect to notice is that the outflow and inflow rates exhibit similar trends but with an offset in time, with depressions in the outflows corresponding to peaks in the inflows, and vice versa. This pattern arises due to the galactic fountain cycle: a certain amount of gas is ejected from the disc as a galactic fountain due to clustered type II SN explosions, creating a peak in the outflow rate. After a time interval of approximately $t_{\text{fount}}$, the fountain gas falls back onto the disc, leading to a peak in the inflow rate. Periods of higher inflow rates result in an increased and more extended SFR, thereby enlarging the SFR scale length. This occurs because gas accretion is not concentrated in the disc inner regions but is rather shifted with respect to the peak of the SFR (see Figure \ref{netrate}). Vice versa, when the outflows are dominant, the SFR scale length diminishes.

 In all the simulations we find that the peaks in the inflow rate roughly coincide with the peaks in gas and stellar scale lengths, we highlight these peaks for \mw \ with dashed vertical lines in the central panel. In contrast, the peaks in the outflow rate correspond to the SFR scale peaks \citep[see also ][for a similar effect on dwarf-mass galaxies]{Zhang2024}. This shows us the intricate connection between outflows/inflows and SFR, which is mediated by stellar feedback and the consequent gas circulation. As discussed earlier, vertical gas inflows are distributed throughout the disc plane, with the majority of the gas accreting at $\approx0.75 R_d$ (Figure \ref{netrate}). This accretion pattern increases gas density in these external regions, leading to an expansion in the size of the gaseous disc and consequently also the stellar disc. The SFR peaks, along with the associated stellar feedback, generate outflows, which are fundamental to the cyclical process of disc growth and evolution.

\section{Comparison with previous works}\label{comparison}

Despite the challenges, gas accretion has been studied both in the Milky Way and in external galaxies. Evidence of cold gas accreting in the Milky Way is observed through intermediate velocity clouds (IVCs, $40<v_{\text{LSR}}<90$ km s$^{-1}$) and high velocity clouds (HVCs, $v_{\text{LSR}}>90$ km s$^{-1}$) \citep[e.g.][]{wakker1997,lehner2010}, gas clouds with anomalous kinematic with respect to the rotation of the disc. HVCs usually have higher velocities, lower metallicities ($0.1<Z/ Z_\odot <1$) and are located farther from the disc (up to 10 kpc) with respect to IVCs, that have higher metallicities ($Z\approx Z_{\odot}$) and are usually close to the disc (within 2 kpc). Historically, these differences have been interpreted as a dichotomy in the origin of these two populations of clouds: the IVCs are produced by the cycle of gas inside galactic fountains and the HVCs are coming from the accretion of gas from the CGM \citep{peck2008}. However, in recent years this picture has been challenged and it has been proposed that HVCs and IVCs are both manifestation of the galactic fountain phenomenon at different velocities \citep{marasco2022,lehner2022}. The IVCs are outflows of gas ejected at lower velocity, therefore they are located closer to the disc and can maintain a metallicity similar to the ISM. HVCs are pushed farther away from the disc and therefore they can mix and decrease their metallicity. 
\citet{zhangzhijie2024} employed the \smuggle \ model in simulations of isolated Milky Way-like galaxies, without the inclusion of a CGM, to study \ion{O}{VI} absorbers. Their findings suggest that key observational properties of low-velocity ($v_{\text{LSR}}\lsim100$ km s$^{-1}$) \ion{O}{VI} absorbers, such as column density and scale height, are well reproduced by the inclusion of galactic fountains driven by stellar feedback. However, the absence of the CGM in their simulations may lead to an underestimation of the column density of high-velocity ($v_{\text{LSR}}\gsim100$ km s$^{-1}$) \ion{O}{VI} absorbers. 

Extrapolating the accretion rate values of observed high- and intermediate-velocity clouds to the whole disc, observations find a maximum accretion rate for HVCs of 0.4 M$_{\odot}$ yr$^{-1}$ \citep{putman2012} and of $1.3-4.3$ M$_{\odot}$ yr$^{-1}$ for IVCs \citep{rohser2016}, which would be enough to sustain the SFR in the Milky Way ($\text{SFR}\approx1.5-3 \ \text{M}_{\odot}$ yr$^{-1}$, \citealt{chomiuk2011, licquia2015, elia2022}). HVCs and IVCs are clearly visible in the gas projections of our simulations (see Figures \ref{metal_edge} and \ref{proj_radial}). Assuming that $40\%$ of the total inflow rate, which averages between 10-20 M$_{\odot}$ yr$^{-1}$ (see Figure \ref{gas_scale}), is attributed to HVCs and IVCs \citep{putman2012, trapp2022}, we find values consistent within a factor of 2 with observed data. It’s important to note that the 9 simulated star-forming galaxies have significantly different SFR (higher than the Milky Way), which will be reflected in the resulting inflow rates.

\citet{li2023} conducted a study on the extraplanar gas of the spiral galaxy NGC 2403, employing a dynamical galactic fountain model. Their findings suggest that fountain-driven accretion can be responsible for an inside-out evolution, effectively cooling the CGM at large radii. In the absence of galactic fountains, cooling would occur predominantly at the centre of the CGM, where gas density is the highest. However, their model applied to observations demonstrates that most gas is accreted at larger radii, a result that aligns closely with the outcomes of our simulations. They observed outflow/inflow rates per unit area similar to those in our work. In their case, the rates peak at the centre. In contrast, our simulations show a central `hole' due to the height at which our rates are computed. \citet{li2023} derived the rates at $z=0$ kpc, a condition that we cannot replicate in our simulations. They derived a coronal gas inflow rate whose shape can be compared to our net vertical inflow rate (Figure \ref{netrate}). We both find a bell-shaped distribution, with the peak skewed towards larger radii. As already mentioned, this is likely due to the increased orbital path of the galactic fountains with radius and to the fact that the SFRD profile is declining with radius.

Different works have focused on the analysis of radial mass flows and their relation to galaxy evolution. Observationally, there is not a clear evidence of the existence of radial mass fluxes strong enough to sustain the formation of new stars in the central regions of disc galaxies. Some references find radial inflows that are comparable with the SFR of the host galaxy, whereas in other works such inflows are not present. For instance, \citet{schmidt2016} studied 10 local spiral galaxies using the \ion{H}{i} Nearby Galaxy Survey. They derived average radial gas mass rates of $\approx 1-3$ M$_{\odot}$ yr$^{-1}$, which will be sufficient to explain the level of SFR in such galaxies. On the other hand, \citet{diteodoro2021} analysed a sample of 54 local spiral galaxies and found small inflow rates with velocities of a few km s$^{-1}$ in only half the galaxies of the sample. They computed an average inflow rate over the entire sample of $(-0.3 \pm 0.9)$ M$_{\odot}$ yr$^{-1}$ at radii larger than the optical disc (inside which most of the star formation occurs). They concluded that the radial inflows are not sufficient to sustain the star formation across the disc and therefore that a different mechanism in addition to secular accretion is necessary. We find that net radial mass rate value, averaged across our 9 simulations, is $\approx 1.72 $ M$_{\odot}$ yr$^{-1}$ with a standard deviation of $\approx 0.86 $ M$_{\odot}$ yr$^{-1}$,  which lies somewhat in between the results of \citet{schmidt2016} and \citet{diteodoro2021}. The net radial mass rates estimated from our numerical experiment can sustain a significant fraction of the SFR observed in simulations. This gas radial motion combined with gas accretion in the central regions of the disc can maintain an active level of star formation in the simulated systems. 

Finally, in recent years, some authors have been using cosmological simulations of Milky Way analogues to investigate vertical and radial gas inflows. For example, \citet{trapp2022} studied gas accretion using FIRE-2 cosmological zoom-in simulations of 4 Milky Way-like galaxies. They found that gas is accreted preferentially in the outskirts of the galaxy, from where the gas moves towards the centre with a mass rate comparable to the galaxy SFR. In their simulations, gas vertical inflows present a bell-shaped distribution and gas particles join the disc at $0.5-1 R_d$, similarly to our results. Radial mass rate and velocities present trends broadly consistent with our findings, but in our case the radial gas motion is not sufficient to entirely sustain the star formation activity of the galaxy. Moreover, \cite{trapp2022} show that most accretion occurs co-rotating and parallel to the disc plane, with only a subdominant contribution from HVCs and IVCs, whereas our results indicate that the majority of the gas is accreted vertically onto the disc, while still having significant radial inflows that move the gas towards the disc centre. However, in \citet{trapp2022} gas accretion is not mainly connected to a tight galactic fountain structure. The differences in the origin of these features are likely due to the different implementations and strengths of stellar feedback in the simulations. In our simulations, using the \smuggle \ model, stellar feedback primarily drives galactic fountains. These fountains result in gas being ejected to moderate heights above the disc and subsequently falling back, often in coherent, vertically aligned trajectories that concentrate near the disc edge. In contrast, the FIRE-2 simulations used in \citet{trapp2022} employ a more energetic stellar feedback model that produces stronger outflows that reach much larger distances into the CGM. As a result, the inflowing gas in their simulations has a more chaotic origin and is not necessarily tied to a galactic fountain cycle.

\section{Summary and Conclusions}\label{summary}

In this work we analysed the baryon cycle in 9 spiral galaxies with different masses and disc sizes, surrounded by a hot CGM, using state-of-the-art $N$-body hydrodynamical simulations. In particular, we focused on the vertical and radial gas mass flows and on their connection to the structure of the disc. We analysed the spatial distribution of outflows (forming the galactic fountain cycle) and inflows linking the vertical accretion and the formation of stars in the centre through radial flows in the plane of the disc. In doing so, we employed the moving mesh code \arepo\ in conjunction with the explicit ISM and stellar feedback model \smuggle. We summarise the main results as follows.

\begin{itemize}

    \item[(i)] We studied inflows and outflows per unit area divided in cold ($T<10^4$ K), warm ($10^4<T<5\times10^5$ K) and hot ($T>5\times10^5$ K) gas. We found that the cold phase at $|z_0|=0.5$ kpc is dominating both gas accretion and ejection followed by the warm and the hot phases. In each simulation we find a similar distribution, with a break point roughly corresponding to the disc size $R_d$, after which the outflow/inflow rates decline rapidly, showing how the outflow/inflow distribution is linked to the structure of the disc.

    \item[(ii)] We found that gas is accreted vertically onto the disc in a bell-shaped distribution, with the peak skewed towards larger radii ($\approx0.75R_d$) compared to the SFR distribution. This offset is due to the orbital paths of galactic fountains and their interaction with the CGM, which increases the accreted material. The hot gas exhibits a flatter distribution and appears to be independent from the warm and cold gas, likely resulting from CGM accretion via a (subdominant in terms of mass) hot cooling flow. In contrast, the cold and warm gas distributions are closely linked. The cold gas is predominantly generated by the galactic fountain cycle, while the warm gas forms at the CGM-disc interface owing to the passage of the galactic fountain clouds, as found in \citet{barbani2023} and confirmed by the simulations in this work.

    \item[(iii)] We examined the radial gas inflows in the disc. We acknowledge the presence of such net inflows with average values $\approx 0.7-3.8$ M$_{\odot}$ yr$^{-1}$ across the 9 simulations. The gas is flowing inward and outwards in a periodic trend with an overall inflowing flux in all the simulations over 2 Gyr. The shape of the radial net inflow distribution presents similarities in all the simulations. The mass rate has its peak at $\approx R_d$, it decreases going inside owing to the higher gas density and to the formation of stars in the central region of the disc. We found that these radial inflow rates have radial velocity $\approx -(3-6)$ km s$^{-1}$ with an almost constant value inside $R_d$. The radial inflows alone are not sufficient to sustain the star formation in the centre, but they can do it in conjunction with vertical inflows which fall at $R<0.4 R_d$, where $90\%$ of the SFR is concentrated. Thus, while the gas accreted from the CGM lands preferentially in the outer regions of the disc, the SFR can be fed by radial motions generated by the angular momentum difference between the CGM and the gas in the disc and by the gas vertically accreted in the centre.

    \item[(iv)] Gas accretion concentrated in the external regions of the disc causes an inside-out evolution scenario, where the galaxy is growing from the inside to the outside. This is confirmed by the slow growth of both the star and gas disc sizes over time in all 9 simulations. In particular, gas and stellar discs grow together with an up-and-down trend linked to stellar evolution. Furthermore, the accretion and ejection of gas are deeply linked to the periodic growth of the disc: peaks in the inflow rate are connected to periods in which the disc is growing, the opposite happens for the outflow rate.

\end{itemize}

The analysis that we have carried out in this work has allowed to link different crucial aspects of gas dynamics in spiral galaxies, collectively known as the baryon cycle. Fresh gas is predominantly accreted from the circumgalactic medium in the outer regions of the disc. This infalling gas has a lower specific angular momentum compared to the gas already present in the disc, leading to an inward flow of gas due to the conservation of angular momentum. At the galaxy centre, where gas density is the highest, a portion of this gas is converted into stars, sustaining the SFR. These newly formed stars impart feedback to the surrounding ISM, eventually driving cold gas clouds out of the disc and forming galactic fountains. As these fountains travel through and interact with the CGM, they create a warm gas phase that eventually cools down and accretes onto the disc at larger radii, thus continuing the baryon cycle. Looking ahead, the inclusion of additional physics, such as magnetic fields, could offer new insights on accretion dynamics. Additionally, studying baryon cycle processes within a full cosmological framework will be crucial for advancing our understanding of spiral galaxies evolution. We defer these aspects to future work.

\begin{acknowledgements}
We acknowledge the CINECA award HP10C9836Y (project IsCb9\_MWOUTF) under the ISCRA initiative, for the availability of high performance computing resources and support. We also acknowledge the use of computational resources from the parallel computing cluster of the Open Physics Hub (\url{https://site.unibo.it/openphysicshub/en}) at the Department of Physics and Astronomy, University of Bologna. RP acknowledges support from PRIN INAF 1.05.01.85.01. FM acknowledges funding from the European Union - NextGenerationEU under the HPC project “National Centre for HPC, Big Data and Quantum Computing” (PNRR - M4C2 - I1.4 - CN00000013 – CUP J33C22001170001). HL is supported by the National Key R\&D Program of China No. 2023YFB3002502, the National Natural Science Foundation of China under No. 12373006, and the China Manned Space Program through its Space Application System. LVS is grateful for partial financial support from NSF-CAREER-1945310 and NSF-AST-2107993 grants.
\end{acknowledgements}

\bibliographystyle{aa}
\bibliography{biblio}

\end{document}